\documentclass[twocolumn]{aastex63}
\usepackage{amsmath}
\usepackage{mathrsfs}
\usepackage{color}
\usepackage{graphicx}
\usepackage[cal=boondox,scr=boondoxo]{mathalfa}

\newcommand{\pressuremodel}{DPP}
\newcommand{\pressuremodelname}{Debiased Pressure Profile}
\submitjournal{ApJ}
\shorttitle{Debiased Galaxy Cluster Pressure Profile}
\shortauthors{He et al.}

\begin{document}

\title{Debiased Galaxy Cluster Pressure Profiles from X-ray Observations and Simulations}

\author{Yizhou He}
\affil{McWilliams Center for Cosmology, Department of Physics, Carnegie Mellon University, Pittsburgh, PA 15213}
\author{Philip Mansfield}
\affil{Department of Astronomy and Astrophysics, University of Chicago, Chicago, IL 60637}
\affil{Kavli Institute for Cosmological Physics, The University of Chicago, Chicago, IL 60637, USA}
\author{Markus Michael Rau}
\affil{McWilliams Center for Cosmology, Department of Physics, Carnegie Mellon University, Pittsburgh, PA 15213}
\author{Hy Trac}
\affil{McWilliams Center for Cosmology, Department of Physics, Carnegie Mellon University, Pittsburgh, PA 15213}
\author{Nicholas Battaglia}
\affil{Department of Astronomy, Cornell University, Ithaca, NY 14853}

\begin{abstract}
We present an updated model for the average cluster pressure profile, adjusted for hydrostatic mass bias by combining results from X-ray observations with cosmological simulations. Our model estimates this bias by fitting a power-law to the relation between the ``true" halo mass and X-ray cluster mass in hydrodynamic simulations (IllustrisTNG, BAHAMAS, and MACSIS). 
As an example application, we consider the REXCESS X-ray cluster sample and the Universal Pressure Profile  (UPP) derived from scaled and stacked pressure profiles. We find adjusted masses, $M_\mathrm{500c},$ that are $\lesssim$15\% higher and scaled pressures $P/P_\mathrm{500c}$ that have $\lesssim$35\% lower normalization than previously inferred. Our \pressuremodelname\ (\pressuremodel) is well-fit by a Generalized Navarro-Frenk-White (GNFW) function, with parameters $[P_0,c_{500},\alpha,\beta,\gamma]=[5.048,1.217,1.192,5.490,0.433]$ and does not require a mass-dependent correction term. When the \pressuremodel{} is used to model the Sunyaev-Zel'dovich (SZ) effect, we find that the integrated Compton $Y-M$ relation has only minor deviations from self-similar scaling. The thermal SZ angular power spectrum is lower in amplitude by approximately 30\%, assuming nominal cosmological parameters (e.g. $\Omega_\text{m}=0.3$, $\sigma_8 = 0.8$), and is broadly consistent with recent Planck results without requiring additional bias corrections.
\end{abstract}
\section{Introduction} \label{sec:intro}

Galaxy clusters are formed by the gravitational collapse of large overdensities and are accompanied by a complex interplay of gravity and baryonic processes. They are ideal probes to study dark energy and the evolution of large scale structure \citep[e.g.][]{2005RvMP...77..207V,2011ARA&A..49..409A}, and their abundance is sensitive to cosmology, meaning that accurate measurements of the cluster mass function and its evolution can provide meaningful cosmological constraints and further our understanding of cosmology in upcoming cluster surveys.

Galaxy clusters have deep gravitational potential wells, and the potential energy of material falling into clusters leads to shock-heating of the gas. This hot, ionized gas emits X-rays through bremsstrahlung radiation, making clusters of galaxies the most common, bright, extended extragalactic X-ray sources. It also makes X-ray observation one of the most attractive methods to detect and characterize galaxy clusters. Due to tight X-ray observable-mass relations, the X-ray temperature $T_\text{X}$, gas mass $M_\text{g}$, $Y_\text{X}=T_\text{X}M_\text{g}$ and X-ray luminosity $L_\text{X}$ inferred from X-ray spectroscopy, have been used as robust mass proxies of galaxy clusters \citep[e.g.][]{2007A&A...474L..37A}. The ACT and the Planck collaborations \citep[e.g.][]{2013JCAP...07..008H, 2016A&A...594A..27P,2018ApJS..235...20H} have been used stacked pressure profiles of the Intracluster Medium (ICM) in galaxy clusters (\citealp{2010A&A...517A..92A}; see also, \citealp{2007ApJ...668....1N}), modeled on X-ray measurements, to interpret survey data of the SZ effect \citep{1970Ap&SS...7....3S}, represented as a distortion in the spectrum of the cosmic microwave background (CMB) due to relic CMB photons inverse Compton scattering off energetic electrons in the galaxy clusters.

When estimating cluster masses from X-ray measurements of density and temperature profiles of the ICM, clusters are generally assumed to be in a dynamical state of hydrostatic equilibrium. However, in the hierarchical structure formation model, galaxy clusters are dynamically active systems and are not in exact hydrostatic equilibrium. Both the latest observations \citep[e.g.][]{2009PASJ...61.1117B,2009MNRAS.395..657G,2009A&A...501..899R,2010PASJ...62..371H,2010ApJ...714..423K,2011MNRAS.414.2101U,2011Sci...331.1576S,2018PASJ...70....9H,Siegel_2018} and numerical simulations \citep[e.g.][]{1990ApJ...363..349E,2004MNRAS.351..237R,0004-637X-705-2-1129,2012ApJ...758...74B,2012ApJ...751..121N,2013ApJ...777..151L,0004-637X-792-1-25,2017MNRAS.469.3069G} find non-thermal gas processes like virialized bulk motions and turbulent gas flows, generated primarily by mergers and accretion during cluster formation, lead to non-trivial pressure support especially in the outskirt of galaxy clusters. Analytical models have also been developed to describe the non-thermal pressure support in intracluster gas and found that it was in excellent agreement with high resolution cosmological hydrodynamic simulations \citep[e.g.][]{doi:10.1093/mnras/stu858,doi:10.1093/mnras/stv036}.
 
Recent work suggests that neglecting the existence of non-thermal pressure in X-ray observations causes systematic underestimation of the hydrostatic masses of galaxy clusters and is a major source of bias in the inferred hydrostatic masses. This is referred to as hydrostatic mass bias. Studies have shown that correcting the absence of non-thermal pressure in hydrostatic equilibrium will help mitigate the tension between cluster mass estimates from weak lensing surveys and from X-ray surface brightness and SZ observations \citep[e.g.][]{doi:10.1093/mnras/stv2504}.

Hydrostatic mass bias has often been assumed to be a constant, parameterized in terms of $b=1-M_{\text{X}/\text{SZ}}/M_\text{WL}$ where $M_{\text{X}/\text{SZ}}$ refers to hydrostatic masses obtained from X-ray or SZ observation and $M_\text{WL}$ refers to results of weak-lensing measurements. Observations giva a range of biases $b=5-30\%$  \citep[e.g.][]{2014MNRAS.443.1973V,2015MNRAS.449..685H,2015AAS...22544304S,10.1093/mnras/stw3322,Battaglia_2016,2016MNRAS.456L..74S,2017A&A...604A..89P,2017MNRAS.472.1946S,2018PASJ...70S..28M}. Numerical simulations \citep[e.g.][]{2007ApJ...655...98N,2012ApJ...758...74B,doi:10.1111/j.1365-2966.2012.20623.x,1367-2630-14-5-055018,2014MNRAS.441.1270L} also point to typical mass biases around $b$=0.20. That hydrostatic bias could depend on cluster mass was not proposed until recently \citep[e.g.][]{2012NJPh...14e5018R}: \citet{doi:10.1093/mnras/stw2899} find that mass bias climbs from 0.20 to 0.40 as cluster masses increase from $M_\text{500c}=10^{14}$ to $10^{15}h^{-1}M_{\odot}$. \citet{2020arXiv200111508B} introduced the Mock-X analysis framework, a multi-wavelength tool that generates synthetic images from cosmological simulations and derives directly
observable and reconstructed properties from these images via observational methods, and applied this framework to explore hydrostatic mass bias for the IllustrisTNG \citep[e.g.][]{Phillepich_et_al_2018,Nelson_et_al_2018,Naiman_et_al_2018,Marinacci_et_al_2018,Springel_et_al_2018}, BAHAMAS \citep{McCarthy_et_al_2017}, and MACSIS \citep{2017MNRAS.465..213B} simulations. They find hydrostatic bias recovered from synthetic X-ray images which shows a significantly stronger mass dependence, increasing from $b=0.0$ at $10^{14}M_\odot$ to $b=0.2$ at $2\times10^{15}M_\odot$. Both studies claim that the key factor causing this mass dependence is the increase in dense, cold gas in cluster outskirts as mass increases. The quadratic dependence of X-ray emission on density causes this cool gas to lower mass estimates for the most massive clusters. Carefully treating hydrostatic mass bias in the recalibration of the ICM pressure models derived from X-ray observation is crucial for better interpreting the angular power spectrum of the thermal SZ signal, reducing systematic uncertainties in cosmological parameters. 

This paper is organized as follows. In Section \ref{sec:methods}, we begin by introducing an analytical approach for correcting hydrostatic mass bias in clusters based on the ``true" simulated mass and the X-ray mass of clusters drawn from the IllustrisTNG, BAHAMAS and MACSIS simulations. We then discuss how to apply this model to the best-fit Generalized Navarro–Frenk–White (GNFW; \citealp{1996MNRAS.278..488Z}) ICM pressure profiles measured in X-ray surveys. In Section \ref{sec:results}, we apply the correction discussed in Section \ref{sec:methods} to the X-ray measurements of cluster masses and the GNFW fit correction to the scaled pressure profiles of the REXCESS cluster sample \citep{2007A&A...469..363B}. We use the corrected characteristic pressures and masses of the REXCESS sample to modify the Universal Pressure Profile (UPP), which gives us a new model for cluster pressures: the Debiased Pressure Profile (\pressuremodel). We use the \pressuremodel{} to study the power-law relation between the integrated Compton parameter and cluster mass. We also calculate the thermal Sunyaev-Zeldovich (tSZ) angular power spectrum with the \pressuremodel{}, and compare with Planck, ACT, and SPT measurements of the tSZ power spectrum. In Section \ref{sec:discussionandconclusions}, we conclude our findings for the mass bias of clusters in the REXCESS sample, the self-similarity of both the new pressure model and the $Y-M$ relation, and the change in amplitude of the tSZ angular power spectrum we get based on the new pressure model. In the end, we also bring up the remaining questions and possible directions for future work. We adopt the following cosmological parameters: $\Omega_m=0.3,\ \Omega_\Lambda=0.7,\ \Omega_b=0.045,\ h=0.7,\ n_s=0.96,\ \sigma_8=0.8$ in this paper.

\section{Methods}\label{sec:methods}

\subsection{ $M_\mathrm{500c}^\mathrm{True}$ v.s. $M_\mathrm{500c}^\mathrm{X-ray}$ of Mock-X}\label{subsec:mass_relation}

\begin{figure*}
\gridline{
    \includegraphics[width=0.48\textwidth]{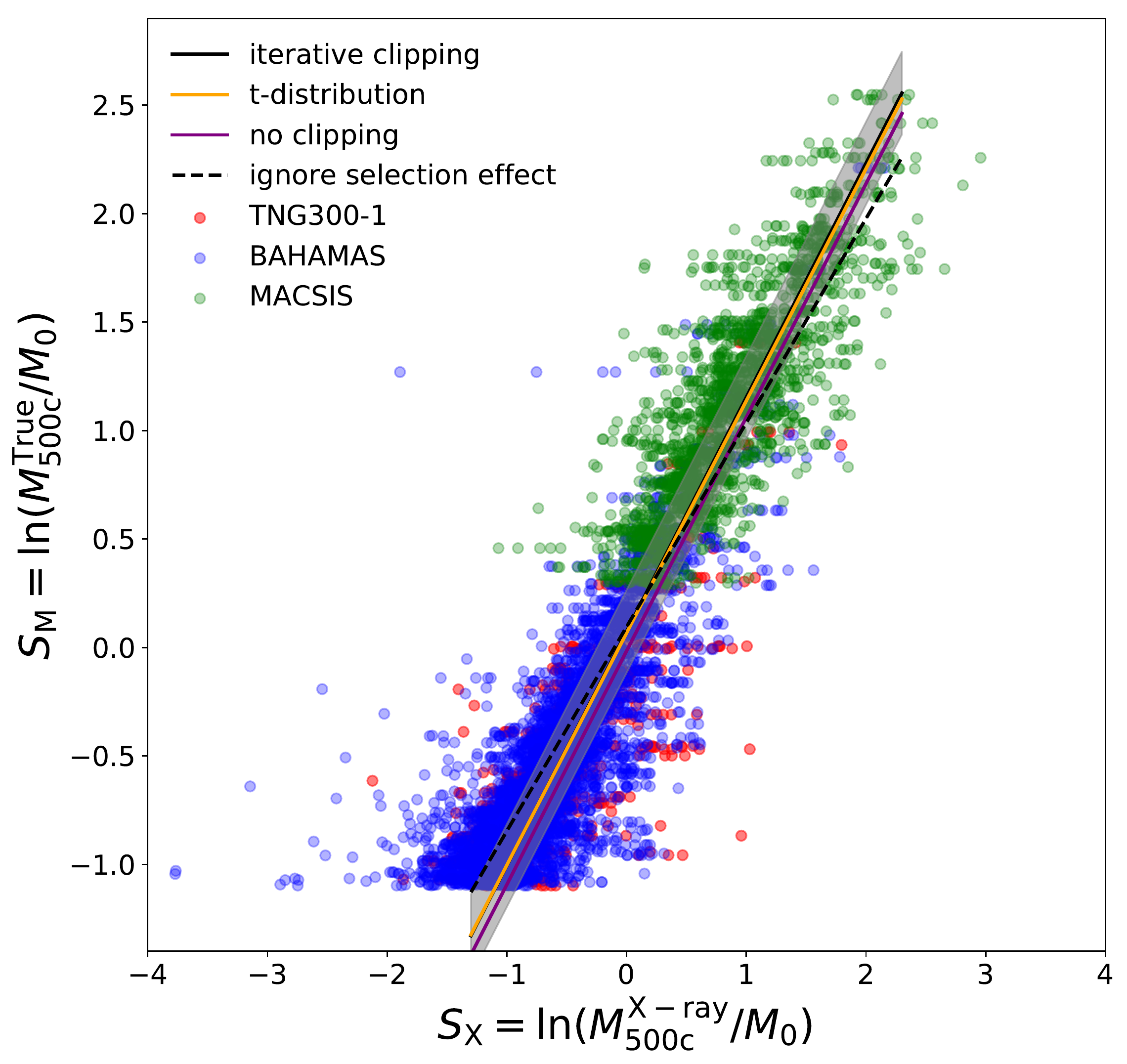}
    \includegraphics[width=0.48\textwidth]{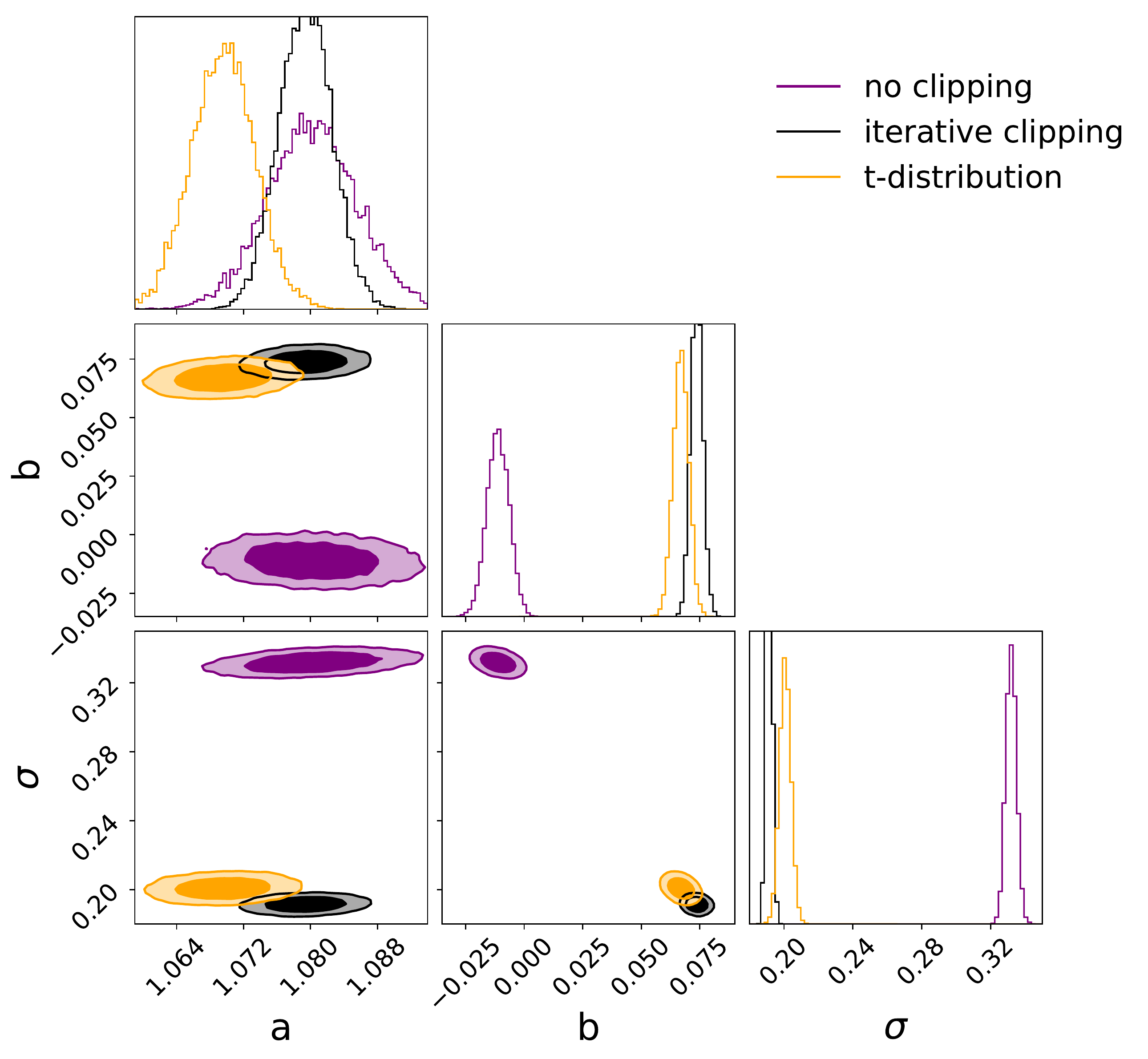}
}
\caption{\textbf{Left}: Normalized ``true'' masses ($M_\text{500c}^\mathrm{True}/M_\text0$) vs.
X-ray masses  ($M_\text{500c}^\mathrm{X-ray}/M_\text0$) for clusters from the IllustrisTNG (red), BAHAMAS (blue), and MACSIS (green) simulations are shown by a scatter plot. The power-law regression described in section \ref{subsec:mass_relation} (solid black line) and the corresponding 68\% scatter (gray shaded region), defined by regression parameter $\sigma$, is plotted over the cluster data. This fiducial approach -- based on iterative clipping -- is consistent with an alternative fit which does not perform clipping but uses a truncated $t$-distribution to account for outliers (solid orange line). Outlier removal has a modest but statistically significant effect on fit results, as shown by a fit which did not account for outliers (solid purple line), but failing to account for mass selection effects (dashed black line) results in a substantially different power-law index. 
\textbf{Right}: Marginalized (1D and 2D) joint posterior probability distributions of the regression model parameters. The dark and light contours show 68\% and 95\% confidence level respectively.}\label{fig:mass_regression}
\end{figure*}

For a spherically symmetrical cluster, hydrostatic equilibrium occurs when the the force of gravity exerted on gas in the cluster is balanced by the gradient of gas pressure:
\begin{equation}\label{eq:HE}
\frac{\mathrm{d}P(r)}{\mathrm{d}r}=-\rho_{\text{gas}}(r)\frac{GM(<r)}{r^2},
\end{equation}
with the gravitational constant, $G$, enclosed mass profile, $M(<r)$, gas pressure profile, $P(r)$, and gas density profile, $\rho_{\text{gas}}(r)$, all with respect to the distance $r$ from the center of the cluster. A combination of X-ray observations like $\text{XMM}-Newton$, CHANDRA and analysis technique taking into account projection and PSF effects have achieved high resolution measurements of the radial electron density profiles, $n_\text{e}(r)$, and the radial temperature profiles, $T(r)$, of galaxy clusters \citep[e.g.][]{2006ApJ...640..691V,2008A&A...487..431C}, which can be used to determine the radial electron pressure profile, $P_\text{e}(r)$, by assuming an ideal gas equation of state, $P_\text{e}(r)=k_\text{B}n_\text{e}(r)T(r)$. Given the electron pressure, the gas thermal pressure $P_\text{th}$ is defined by $P_\text{th}(r)=P_\text{e}(r)\mu_\text{e}/\mu $ where $\mu$ is the mean mass per gas particle, and $\mu_\text{e}$ is the mean mass per electron.

In addition to the thermal motion of the gas, other sources of gas pressure - including viralized bulk motion, turbulence, cosmic rays, and magnetic fields - also provide non-trivial pressure support \citep[e.g.][]{1997ApJ...477..560E,2008MNRAS.388.1062C,2015ASSL..407..599B}. For realistic equilibrium systems, the gas pressure, $P$, in Eq.~\ref{eq:HE} is replaced by $P=P_\text{th}+P_\text{nth}$, where $P_\text{nth}$ refers to any non-thermal pressure acting on the intracluster gas. X-ray-measured cluster masses are derived from the assumption of hydrostatic equilibrium with only thermal gas pressure, which means that the contribution of non-thermal pressure can cause X-ray measurements to underestimate cluster masses systematically.

Numerical simulations provide a vital resource for characterizing the mass bias as the properties of simulated galaxy clusters are known exactly. \citet{2020arXiv200111508B} developed the Mock-X analysis framework, which can generate synthetic X-ray images and derives halo properties (e.g. gas density and temperature profiles) via observational methods, which can be used to derive hydrostatic mass in mock X-ray observations. Hydrostatic mass bias is equal to the ratio of the hydrostatic mass to the ``true'' (overdensity) mass of simulated clusters identified through SUBFIND \citep[e.g.][]{2001MNRAS.328..726S,2009MNRAS.399..497D} in simulations. Studies \citep[e.g.][]{doi:10.1093/mnras/stw2899,2020arXiv200111508B} also point out that the bias of hydrostatic mass estimated with density and temperature profiles derived from the spectroscopic analysis show a much stronger mass-dependence than those estimated from the true mass-weighted temperature profiles. These simulated spectroscopic temperatures emulate the observational procedure for measuring X-ray temperatures and thus we compare against them in this analysis.

A number of the aforementioned numerical studies have measured $M_\text{500c}^\mathrm{X-ray}$/$M_\text{500c}^\mathrm{True}.$ However, observationally, one only has access to $M_\text{500c}^\mathrm{X-ray}.$ This means that we must invert these relations to give $M_\text{500c}^\mathrm{True}/M_\text{500c}^\mathrm{X-ray}$ as a function of $M_\text{500c}^\mathrm{X-ray}$ (a deceptively complex task). Here, $M_{\rm 500c}$ is the ``overdensity mass'' and corresponds to the mass within a spherical boundary which has an average density equal to 500 times the critical density, $\rho_{\rm crit}.$

In this work, we present an efficient approach to estimate the true cluster mass by utilizing both the X-ray and ``true" masses of simulated clusters, $M_\text{500c}^\mathrm{True}$ from \cite{2020arXiv200111508B}. We adopt a power-law model for the scaling relation between $M_\text{500c}^\mathrm{True}$ and $M_\text{500c}^\mathrm{X-ray}$. This is a linear model in logarithmic scale. For convenience, we denote
\begin{align}
    S_\text{X} & = \ln(M_\mathrm{500c}^\mathrm{X-ray}/M_0), \notag\\
    S_\text{M} & = \ln(M_\mathrm{500c}^\mathrm{True}/M_0),
\end{align}
with $M_\text0=3\times10^{14}M_\odot$. For fixed $S_\text{X}$, we assume a linear relation between $S_\text{M}$ and $S_\text{X}$ where the error, $\epsilon,$ follows a Gaussian distribution:
\begin{align}
    \label{eq:fit_model}
    S_\text{M} & = aS_\text{X}+b+\epsilon,\\
    \epsilon & \sim \text{Norm}(0,\sigma^2),
\end{align}
where $a,b,\sigma$ are free parameters. 

We notice that clusters drawn from simulations are selected in terms of a certain mass threshold, which means we also need to consider this selection effect in our model when fitted to simulation data. For given $S_\text{X}$ and $S_\text{M}$ of a simulated cluster, we use a truncated normal distribution to model the likelihood 
\begin{equation}
    p(S_\text{M}|S_\text{X},S_\text{T},\vec{\theta})=\frac{A(S_\text{X},S_\text{T},\theta)}{\sqrt{2\pi\sigma^2}}\exp\left[\frac{(aS_\text{X}+b-S_\text{M})^2}{2\sigma^2}\right],
\end{equation}
where $\vec{\theta}=(a,b,\sigma)$ denotes the free parameters. $S_\text{T}$ is the truncation parameter defined by $S_\text{T}=\log(M_\text{500c}^\mathrm{T}/M_\text0)$ and $M_\text{500c}^\mathrm{T}$ is the mass threshold for a given simulated cluster sample. $A(S_\text{T},S_\text{X},\vec{\theta})$ is the normalization factor for a normal distribution, $\text{Norm}(aS_\text{X}+b,\sigma^2)$, truncated with a lower bound $S_\text{T}$:
\begin{equation}
    A(S_\text{T},S_\text{M},\vec{\theta})=\left[1-\Phi\left(\frac{S_\text{T}-aS_\text{X}-b}{\sigma}\right)\right]^{-1},
\end{equation}
where $\Phi$ denotes the cumulative distribution function (CDF) of standard normal. We set
\begin{align}
    a,b&\sim \mathcal{U}(-5,5)\\
    \sigma&\sim \mathcal{U}(0,5)
\end{align}
as priors, where $\mathcal{U}$ denotes the uniform distribution. We can then write out the posterior for the parameters
\begin{align}
    p(\vec{\theta}|D)\propto p(a)p(b)p(\sigma)\prod\limits_{\alpha}\left[\prod\limits_{i=1}^{N_\alpha}p(S_{\text{M},\alpha}^i|S_{\text{X},\alpha}^i,S_{\text{T},\alpha},\vec{\theta})\right],
\end{align}
where $D=\{D_1,D_2,D_3\}$ is a data vector of log-scaled masses of simulated cluster sample drawn from IllustrisTNG, BAHAMAS and MACSIS simulations denoted by $\alpha=1,2,3$, and $D_\alpha=\{(S_{\text{M},\alpha}^i,S_{\text{X},\alpha}^i),i=1,\dots\,N_\alpha\}$. Each simulation uses a different $S_{\text{T},\alpha}$. Mass thresholds, $M_\text{500c}^\text{T},$ are set to be $10^{14}M_\odot$ for IllustrisTNG and BAHAMAS, and $4\times10^{14}M_\odot$ for MACSIS. Details about these simulations can be found in \citet{2020arXiv200111508B}.  

We note that the IllustrisTNG, BAHAMAS, and MACSIS simulations adopt different numerical methods
or subgrid physics, which may introduce differences in the derived cluster profiles and systematics in the mass estimation of the mock X-ray observation. This will be accounted in the intrinsic scatter in our regression model for the relation of $M_\text{500c}^\mathrm{True}$ v.s. $M_\text{500c}^\mathrm{X-ray}$ since the fit is performed on all simulations simultaneously. 

We explore the parameter space by Markov Chain Monte Carlo (MCMC), using \texttt{emcee} \citep{2013PASP..125..306F} for the sampling. We discard the initial steps suggested by the integrated autocorrelation time \citep{2019JOSS....4.1864F}, which estimate the number of steps that are needed before the chain “forgets” where it started. This step ensures the samples well "burnt-in". Regression results for the linear model and uncertainty are reported in Table~\ref{tab:reg_params}.

A small fraction, $\sim$9\%, of simulated clusters have abnormally high or abnormally low $M_\text{500c}^\mathrm{True}$/$M_\text{500c}^\mathrm{X-ray}$, suggesting ratios outside the observed range \citep[e.g.][]{2019ApJ...875...63M}.
Most cases appear in low-mass clusters, and may be due to the numerical noise when resolving the X-ray mass of simulated clusters from synthetic images. The steep slope of the mass function causes these unreliable low-mass data points to significantly influence the mean $M_\text{500c}^\mathrm{True}$ at high $M_\text{500c}^\mathrm{X-ray}$. In addition to the numerical noise, merging events, or certain AGN activities in the unrelaxed clusters could lead to a less spherical cluster. The thermodynamic profiles and corresponding X-ray mass of these less regular clusters will be recovered with more systematic uncertainty because clusters' profiles and masses are derived assuming spherical symmetry in the Mock-X analysis \citep{2020arXiv200111508B}, which could also result in an extreme value of $M_\text{500c}^\mathrm{True}$/$M_\text{500c}^\mathrm{X-ray}$. For a more concrete conclusion, detailed studies are required for these peculiar cluster samples with abnormal values of $M_\text{500c}^\mathrm{True}$/$M_\text{500c}^\mathrm{X-ray}$ in the future work.

To mitigate the effect introduced by the simulated clusters with extreme values of $M_\text{500c}^\mathrm{True}$/$M_\text{500c}^\mathrm{X-ray}$, we iteratively remove outlier clusters falling outside the 2$\sigma$ region of the regression results until the prediction for $M_{500c}^\mathrm{True}$ derived from the linear model for $M_{500c}^\mathrm{True}$ v.s. $M_{500c}^\mathrm{X-ray}$ converges to 1\% agreement with the previous iteration. This is performed for clusters within the X-ray mass range $M_{500c}^\mathrm{X-ray}=10^{14}-10^{15}M_\odot$. To test the impact of this method for removing outliers, we also used a truncated $t$-distribution \citep[e.g.][]{10.1093/biomet/83.4.891} to model the uncertainty, $\epsilon$, which is another approach to alleviate the effect of outlier samples. 

In Figure~\ref{fig:mass_regression}, we plot $S_\text{M}$ v.s. $S_\text{X}$ for the IllustrisTNG, BAHAMAS, and MACSIS cluster samples. We also show the regression results for the linear relation between log-scaled ``true'' and X-ray masses, considering both truncation effects and the influence of outlier clusters. The intrinsic scatter in our linear model is determined by the parameter $\sigma$. We find the slope parameter $a=1.079$ is greater than 1, which indicates the ratio of $M_\text{500c}^\mathrm{True}$ to $M_\text{500c}^\mathrm{X-ray}$ is mass-dependent, and hydrostatic mass bias increases with cluster mass. We also plot the regression results for the linear relation by modeling the uncertainty, $\epsilon,$ with a truncated $t$-distribution. For comparison, we also show regression results without removing outlier clusters and without modeling truncation effects. 

The alternative fit for $M_\text{500c}^\mathrm{True}$ v.s. $M_\text{500c}^\mathrm{X-ray}$, which does not perform clipping but uses a truncated $t$-distribution to account for outliers, is in good agreement with the results of iterative 2$\sigma$ clipping methods. Regression results of the two methods find similar values for the slope parameter and the discrepancy between $M_\text{500c}^\mathrm{True}$ for $10^{14}M_\odot<M_\text{500c}^\mathrm{X-ray}<10^{15}M_\odot$ is at the $\sim1\%$ level. Comparing with the fit which did not account for outliers, we find outlier removal has a modest but statistically significant effect on fit results. We also find failing to account for mass selection effects results in a bad fit to the simulation data and a substantially different power-law index.

\begin{table}
\begin{center}
\begin{tabular}
{ p{2.0cm}||p{1.6cm}|p{1.6cm}|p{1.6cm} }
 \hline
 Model parameters &$a$ &$b$ &$\sigma$ \\
 
 \hline
 iterative clipping &1.079 $\pm$0.003 &0.074 $\pm$0.002 &0.191 $\pm$0.001 \\
 \hline
  $t$-distribution &1.070 $\pm$0.004 &0.067 $\pm$0.003 &0.201 $\pm$0.002 \\
 \hline
  no clipping &1.080 $\pm$0.005 &-0.011 $\pm$0.005 &0.332 $\pm$0.003 \\
 \hline
\end{tabular}\caption{Best-fitting parameters for Eq.~\ref{eq:fit_model} for the cluster data from the IllustrisTNG, BAHAMAS, and MAC-SIS simulations. Each row shows a different method for accounting for outlier clusters.}\label{tab:reg_params}
\end{center}
\end{table}

\subsection{Hydrostatic Bias for Pressure Models}\label{subsec:HSBforP}

When we fit an analytical model like a GNFW profile to the radial pressure profile of a galaxy cluster, a common approach taken is to normalize the pressure and radius by the characteristic pressure $P_{500c}$ and radius $R_{500c}$, both of which can be directly computed at a given cluster mass. 
If the mass of galaxy clusters in X-ray measurements suffers from hydrostatic bias, the characteristic pressure and radius will as well. For convenience, we define a new variable for the hydrostatic mass bias, 
\begin{equation}
    B_\text{M}=M_\text{500c}^\mathrm{True}/M_\text{500c}^\mathrm{X-ray},
\end{equation}
then the radius bias, $B_\text{R},$ and pressure bias, $B_\text{P},$ can be obtained from scaling relations, although the latter relies on the assumption of a specific model for the pressure profile. For a spherical cluster, $R_\text{500c}\propto M_\text{500c}^{1/3}$, so hydrostatic bias for cluster radius is defined by
\begin{equation}\label{eq:15}
B_\text{R}=B_\text{M}^{1/3}.
\end{equation}
If we assume that pressure follows a GNFW profile given by $P(r)=P_\text{500c}(M_\text{500c},z)\mathbb{P}(x)$ \citep{2007ApJ...668....1N}, where $x=r/R_\text{500c}$ and
\begin{align}\label{eq:P500c}
P_\text{500c}(M_\text{500c},z)=&1.65\times 10^{-3}h(z)^{8/3} \nonumber\\
&\times\left[\frac{M_{500c}}{3\times10^{14}M_{\odot}}\right]^{2/3}h_{70}^2\,\mathrm{keV\,cm^{-3}},
\end{align}
$\mathbb{P}(x)$ is the scaled profile, with the form
\begin{equation}\label{eq:16}
\mathbb{P}(x)=\frac{P_0}{(c_{500}x)^{\gamma}\left[1+(c_{500}x)^{\alpha}\right]^{(\beta-\gamma)/\alpha}} ,
\end{equation}
where $c_{500}$ is the concentration, $P_\text{0}$ is the normalization parameter, and the parameters $\alpha,\,\beta,\,\gamma$ determine the power-law slopes of different region of the cluster. Since $P_\text{500c}\propto M_\text{500c}^{2/3}$ according to Eq.~\ref{eq:P500c}, the pressure bias is
\begin{equation}\label{eq:17}
B_\text{P}=B_\text{M}^{2/3} .
\end{equation}
The bias parameters $B_\text{R}$ and $B_\text{P}$ can be used to debias GNFW fits to X-ray measurements of thermal pressure profiles by rescaling $c_{500}$ and $P_{0}$ with the following bias correction factors:
\begin{equation}\label{eq:debias_c500}
c_{500} = c_{500}^{\text{bias}}\times B_\text{R},
\end{equation}
\begin{equation}\label{eq:debias_P0}
P_{0} = P_{0}^{\text{bias}}/B_\text{P}.
\end{equation}
We note that radius and pressure biases have a one-to-one relation with $B_\text{M}$, so uncertainty in $B_\text{M}$ can be converted to $B_\text{R}$ and $B_\text{P}$ by
\begin{equation}
    \sigma_{\ln B_{\rm{R}}}=\sigma_{\ln B_{\rm{M}}}/3,\sigma_{\ln B_{\rm{P}}}=2\sigma_{\ln B_{\rm{M}}}/3.
\end{equation}

\section{Results}\label{sec:results}

\subsection{Mass Adjustment of the REXCESS Sample}\label{subsec:rexcess mass correction}

We apply our linear model for $S_\text{M}$ vs. $S_\text{X}$ to the hydrostatic X-ray masses of the REXCESS cluster sample to estimate the true masses of these clusters. REXCESS is a representative sample of local clusters at redshifts $0.0<z<0.2$ which spans a mass range of $10^{14}M_\odot<M_{500c}<10^{15}M_\odot$ \citep{2010A&A...517A..92A}. REXCESS clusters are drawn from the $\mathsf{REFLEX}$ catalog and were studied in-depth by the $XMM-Newton$ Large Programme. A description of the $\mathsf{REFLEX}$ sample and of $XMM-Newton$ observation details can be found in \citet{2007A&A...469..363B}. We correct the hydrostatic mass of 31 local clusters from the REXCESS sample measured by X-ray observation,
\begin{equation}
    M_\text{500c}^{\rm{True}}/M_\text0=e^b\times(M_\text{500c}^{\rm{X-ray}}/M_\text0)^a,
\end{equation}
where $a$ and $b$ are the regression parameters for the linear model reported in Table \ref{tab:reg_params}. We find that the X-ray measured hydrostatic masses of clusters in the REXCESS sample are underestimated by approximately 7\% on average. The bias climbs from 0$\%$ to 15$\%$ as cluster X-ray mass increases from $10^{14}M_\odot$ to $10^{15}M_\odot$.
 
The regression parameter $\sigma$ is the intrinsic scatter in $S_\text{M}=\ln(M_\text{500c}^\mathrm{True}/M_0)$ and can be used to characterize the uncertainty in the corrected mass of REXCESS clusters at a given predicted $M_\text{500c}^\mathrm{True}:$
\begin{align}
\label{eq:M_sigma}
\sigma_{M_\text{500c}^\mathrm{True}}\simeq M_\text{500c}^\mathrm{True}\sigma.
\end{align}
With the first order approximation, this scatter yields significant uncertainties for individual objects, around $\approx$20\%, for corrected cluster masses. $\sigma$ also defines scatter in the mass bias $B_\mathrm{M}$, radius bias $B_\mathrm{R}$, and pressure bias $B_\mathrm{P}$ in log scale:
\begin{equation}
    \sigma_{\ln B_{\rm{M}}}=\sigma,\sigma_{\ln B_{\rm{R}}}=\sigma/3,\sigma_{\ln B_{\rm{P}}}=2\sigma/3.
\end{equation}
These allow us to estimate the modeling uncertainty in the debiased pressure, radius and mass.

\subsection{Adjustment of the Universal Pressure Profile}\label{subsec:mupp}

\begin{figure}
\includegraphics[width=\hsize]{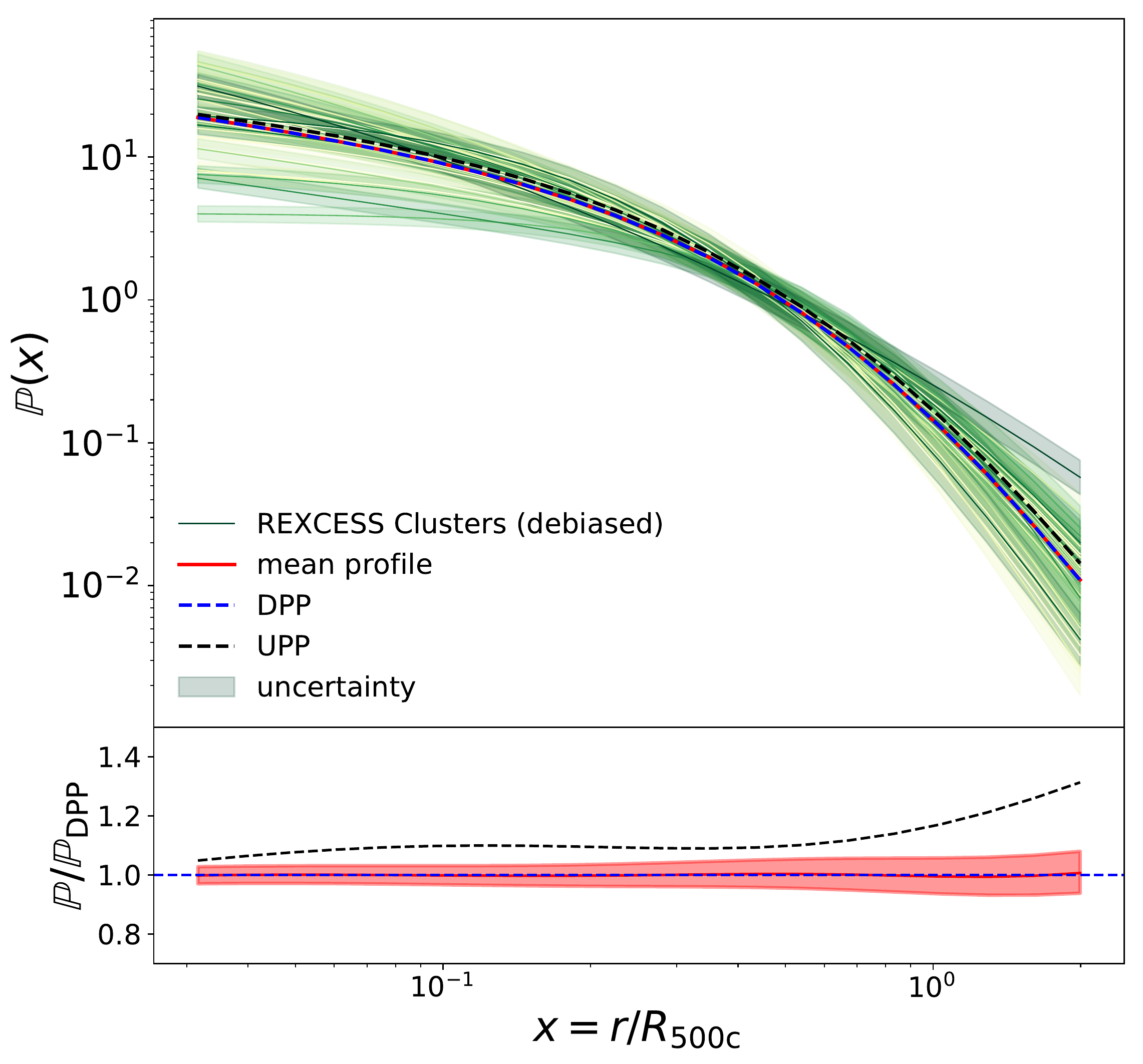}
\caption{\textbf{Top}: Individual GNFW fits for the scaled pressure profiles of each cluster in REXCESS sample after $R_{\rm 500c}$ and $P_{\rm 500c}$ have been corrected for hydrostatic mass bias (solid green lines) with uncertainty estimated from the scatter in the mass bias (green semitransparent bands). Also shown are the mean profile (dashed blue line) of the corrected samples and the best-fitting GNFW profile to the median, $\mathbb{P}(x)$ (solid red line). The best-fitting $\mathbb{P}(x)$ of the uncorrected UPP model (dashed black line) is also plotted for comparison. \textbf{Bottom}: The ratio between $\mathbb{P}(x)$ of the UPP model (dashed black line) and the mean corrected profile of the REXCESS sample (solid red line) with respect to the corrected $\mathbb{P}(x)$ (dashed blue line) are shown. Uncertainty in the adjusted mean pressure profile (red semitransparent band) is calculated through the procedure discussed in Section \ref{subsec:mupp}.}\label{fig:pressure model}
\end{figure}

The Universal Pressure Profile (UPP) is a model for ICM thermal pressure profiles developed by \citet{2010A&A...517A..92A} which was calibrated off the REXCESS sample. For each cluster in the sample, the pressure profile -- derived along with the X-ray measurements of gas density and temperature profiles -- is scaled with the characteristic pressure $P_\text{500c}$ and cluster radius $R_\text{500c}.$ As discussed in Section \ref{subsec:HSBforP}, both $R_{\rm 500c}$ and $P_{\rm 500c}$ are dimensional rescalings of $M_{\rm 500c},$ which itself is measured from a $M_\text{500c}-Y_\text{X}$ relation \citep{2006ApJ...650..128K,2007ApJ...668....1N,2007A&A...474L..37A} which was calibrated on biased hydrostatic mass estimates.
Note that \citet{2007A&A...474L..37A} itself does not use the REXCESS sample, which could potentially allow selection bias to creep in.
The UPP model is widely used for characterizing cluster masses in SZ surveys \citep[e.g.][]{2013JCAP...07..008H,2016A&A...594A..27P,2018ApJS..235...20H} and is expressed as  
\begin{align}\label{eq:upp}
P(x,M_\text{500c},z)=&P_\text{500c}(M_\text{500c},z)\nonumber\\
&\times\mathbb{P}(x)\left[\frac{M_\text{500c}}{3\times10^{14}h_{70}^{-1}M_{\odot}}\right]^{\alpha_\text{P}(x)},
\end{align}
with variables taking the same meaning as in Section \ref{subsec:HSBforP}. The empirical term, $(M_\text{500c}/3\times10^{14}h_{70}^{-1}M_{\odot})^{\alpha_\text{P}(x)}$, reflects the deviation from standard self-similar scaling with $\alpha_\text{P}(x)=0.22/(1+8x^3)$. A GNFW profile, $\mathbb{P}(x)$, is fit against the (geometric) mean profile of the scaled REXCESS sample.

The hydrostatic bias that we found for $M_\text{500c}$ in the REXCESS sample is transferred to the normalization of observed pressure profiles through the resultant changes in $P_\text{500c}$ and $R_\text{500c}.$ For each REXCESS cluster, we use the GNFW pressure profile provided in \citet{2010A&A...517A..92A} and rescale $P_\text0$, and $c_\text{500}$ according to Eq.~\ref{eq:debias_c500} and \ref{eq:debias_P0} to get the debiased fits for each cluster. We then evaluate the geometric mean of the scaled profiles, $P_\mathrm{m}$, and fit it with a GNFW model in the log-log plane. We also estimate the uncertainty in the mean profile by approximating the uncertainty in each corrected pressure profile via lognormal distributions with variances $\sigma_{\ln{R}}$ and $\sigma_{\ln{P}}$. Moreover, we use this uncertainty to define the 68$\%$ range for the mean profile confined by a high profile, $P_\mathrm{h}$, and a low profile, $P_\mathrm{l}$. 

We fit new GNFW models to the mean, high and low profiles discussed above, fixing the outer slope parameter to $\beta= 5.490$ as was done in the original UPP model. In \citet{2010A&A...517A..92A}, the GNFW model of the UPP is fitted to the observed average scaled profile in the radial range [0.03–1]$R_\mathrm{500c}$, combined with the average simulation profile beyond $R_\mathrm{500c}$ which is crucial for determining the outer slope $\beta$. In our paper, the GNFW model is fitted to the debiased observed profiles within $R_\mathrm{500c}$, but we lack information beyond this radius. So we choose to keep $\beta$ as same as its original value in the UPP model. The best fitting parameters of the GNFW models for $P_\mathrm{m}$, $P_\mathrm{h}$, and $P_\mathrm{l}$ are reported in Table~\ref{tab:pressure_params}.

\begin{table}
\begin{center}
\begin{tabular}
{ p{1.5cm}||p{1.2cm}|p{1.2cm}|p{1.2cm}|p{1.2cm} }
 \hline
 GNFW parameters &$P_\text0$ &$c_\text{500}$ &$\alpha$ &$\gamma$\\
 \hline
 UPP       &8.403 &1.177 &1.051 &0.3081 \\
 \hline
 $P_\mathrm{m}$ &5.048 &1.217 &1.192 &0.433 \\
 \hline
 $P_\mathrm{h}$ &5.159 &1.204 &1.193 &0.433 \\
 \hline
 $P_\mathrm{l}$ &4.939 &1.232 &1.192 &0.432 \\
 \hline
\end{tabular}\caption{Parameters for GNFW fits to the mean ($P_{\rm m}$), high ($P_{\rm h};$ +1$\sigma$), and low ($P_{\rm l};$ -1$\sigma$) profiles, as well as parameters for the dimensionless pressure profile of the UPP model.}
\label{tab:pressure_params}
\end{center}
\end{table}

In the top panel of Figure~\ref{fig:pressure model}, we plot corrected GNFW fits to the debiased pressure profiles for each of the 31 REXCESS clusters. As discussed in Section \ref{subsec:rexcess mass correction}, the scatter in $B_\text{M}$ is significant and introduces non-negligible uncertainty to the debiased pressure profiles of the REXCESS sample. We also show the uncertainty in the debiased pressure profile for each RECXESS cluster considering the uncertainty as determined by $\sigma_{\ln B_{\text{R}}}$ and $\sigma_{\ln B_{\text{P}}}$. We show the geometric mean of these scaled profiles, the fit to this curve, and the UPP model for comparison. The dispersion in these scaled pressure profiles is significant in the core both before and after debiasing regions due to the various dynamical states, including both the cool core and morphologically disturbed clusters of the REXCESS sample \citep{2010A&A...517A..92A}. The mean of the debiased scaled pressure profiles and its GNFW fit, $\mathbb{P}(x)$, is lower than in the original UPP model. In the bottom panel of Figure~\ref{fig:pressure model}, we plot the fractional difference between both the UPP and the debiased mean scaled profile against our best fit to the mean scaled profile. We also show uncertainty in the pressure model with the red semitransparent region.

The UPP is $\approx5\%$ higher than the mean of the debiased pressure profile in the center of the cluster and gradually climbs to $20\%$ at $R_\text{500c}$, and reaches almost $\approx35\%$ at the outermost outskirts. Only weak scattering is found for the adjusted scaled pressure profile compared to the uncertainty of scaled pressure profile of each REXCESS cluster,  which is due to the assumption of using Gaussian approximating the uncertainty of individual profile in logarithmic scale at fixed radii, and uncertainty of the mean decreases with the growth of the sample size of the REXCESS clusters.  

\begin{figure}
\includegraphics[width=\hsize]{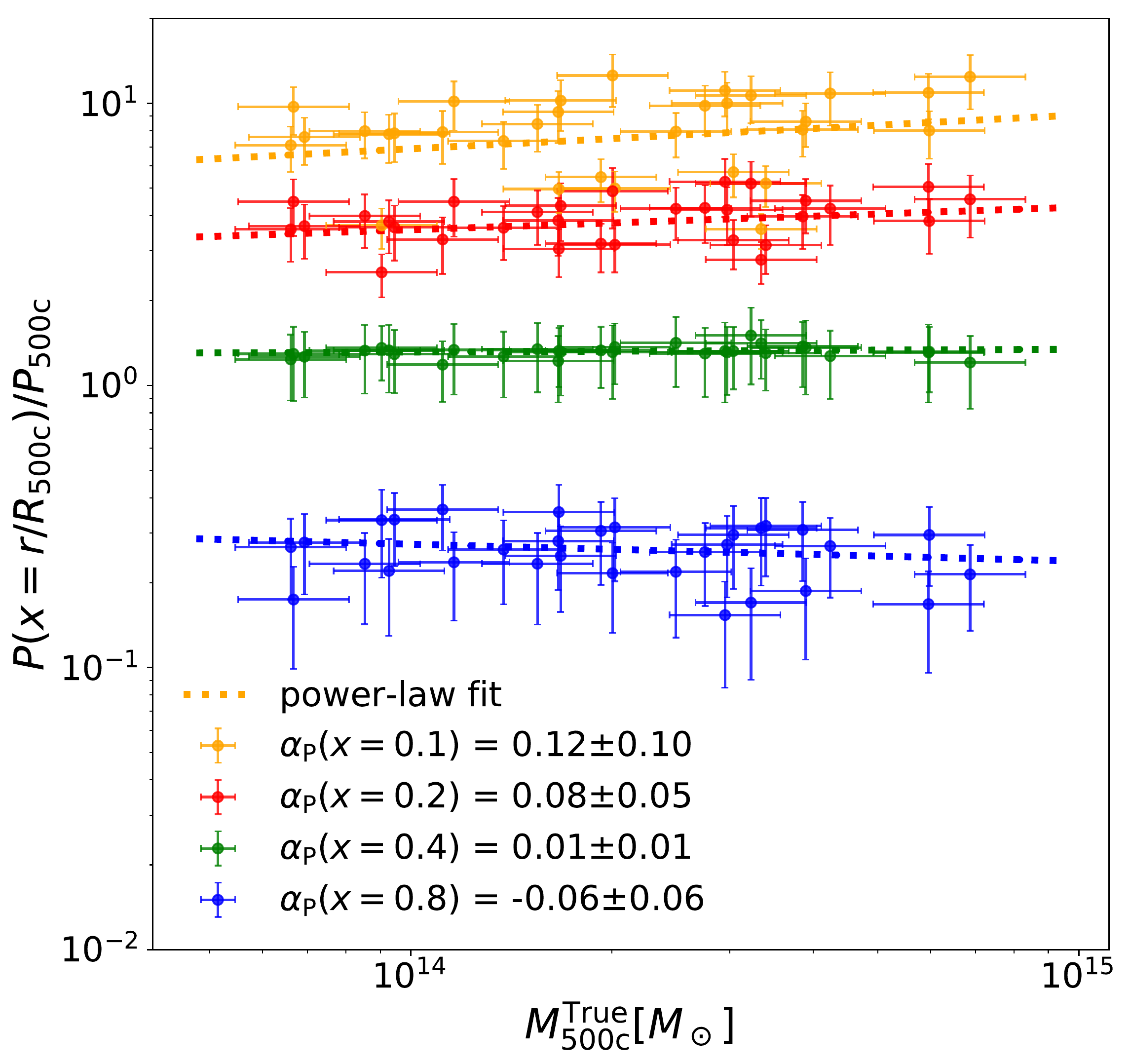}
\caption{Deviations from self-similarity as a function of mass and radius. Debiased pressure is plotted against corrected $M_\text{500c}$ at different scaled radii $x=r/R_\text{500c}$: 0.1 (red), 0.2 (orange), 0.4 (green) and 0.8 (blue). The pressure implied by the best-fitting GNFW pressure profiles at these radii for the 31 clusters in the REXCESS sample are shown as points. We fit power-laws for each value of $x$ (dashed lines) to determine the mass dependence of cluster pressures. Error bars show the uncertainty introduced by the scatter in $B_\text{M}$ while correcting cluster masses and recalibrating the GNFW fit of each RECXESS cluster. After debiasing the pressure profiles, we find no evidence for deviations from self-similarity.}\label{fig:self-similarity deviation} 
\end{figure}

\subsection{Self-similarity of the Pressure Profile}\label{subsec:self-similarity for pressure}

We also explore whether the REXCESS pressure profiles deviate from self-similarity by studying their radial variation as a function of mass. To do this, we look for mass trends in $P(x)/P_\text{500c}$ in our debiased profiles. We evaluate these profiles at $x=r/R_\text{500c}=$ 0.1, 0.2, 0.4, 0.8. This range of radii avoids either too small or too large values of $x$. We avoid larger scaled radii because X-ray measurements pressure profiles in REXCESS clusters rarely get beyond $R_{500c}$ \citep{2010A&A...517A..92A}.\footnote{Also note that the $R_{\rm 500c}$ values in \citep{2010A&A...517A..92A} are biased low by $R_\text{500c}^{\mathrm{X-ray}}/R_\text{500c}^{\mathrm{True}}\sim 0.95$, meaning that the profiles extend to smaller radii than reported in the original paper.}  We avoid taking a smaller value of $x$ because the REXCESS sample contains systems with various dynamical states which can alter the state of gas in the center of the cluster. Following \citet{2010A&A...517A..92A}, we fit a power-law of the form $P/P_\text{500c}\propto M_\text{500c}^{\alpha_\text{P}(x)}$ to each set of points weighted by uncertainties on both cluster masses and pressure following the orthogonal regression approach, proposed for the analysis when both the dependent and the independent variables are random. Best fitting results are represented in Table~\ref{tab:slope_params}.

In Figure~\ref{fig:self-similarity deviation}, we show the results of this fit, with different colors representing different scaled radii and error bars representing uncertainty due to the intrinsic scatter in $B_\text{M}$. We show the best-fit power-laws to each set of points and the values of their power indices.

\begin{table}
\begin{center}
\begin{tabular}
{ p{1.5cm}||p{1.2cm}|p{1.2cm}|p{1.2cm}|p{1.2cm} }
 \hline
 $\alpha_\mathrm{P}(x)$ &x=0.1 &x=0.2 &x=0.4 &x=0.8\\
 \hline
 UPP       &0.22 &0.21 &0.15 &0.06\\
 \hline
 \pressuremodel &0.12\ $\pm$0.10 &0.08\ $\pm$0.05 &0.01\ $\pm$0.01 &-0.06\ $\pm$0.06\\
 \hline
\end{tabular}\caption{Comparison of the best-fitting  $\alpha_\text{P}(x)$ in the UPP and \pressuremodel\ models. Note that under the \pressuremodel{} model, $\alpha_{\rm P}(x)$ is consistent with zero at all radii.}\label{tab:slope_params}
\end{center}
\end{table}

Our study of the debiased scaled pressure profiles of the REXCESS cluster sample finds that $\alpha_\text{P}(x)$ at all radii are consistent with zero, which means a less significant deviation from standard self-similarity compare to the UPP model. We can observe a radial dependence of $\alpha_\text{P}(x)$ similar to that found in the UPP model. However, this term in UPP is treated as a second-order deviation term in addition to a constant modification of the standard self-similarity, $\alpha_\text{P}\sim 0.12$, which can be neglected in first-order approximation. Based on the discussion above, we see no evidence for deviations from self-similarity, which would require the mass-dependent term in Eq.~\ref{eq:upp}. We modify the UPP by eliminating the deviation term and get a simplified model for ICM pressure profiles, \pressuremodelname\ (\pressuremodel) :
\begin{align}\label{eq:22} 
P_\text{\pressuremodel}(x,M_\text{500c},z)=&P_\text{500c}(M_\text{500c},z)\times\mathbb{P}(x).
\end{align}
Here, parameters take on the same meaning as in $P_\mathrm{m}$ in Table~\ref{tab:pressure_params}.

\subsection{$Y-M$ Relation}\label{subsec:Y-M}

\begin{figure}
\includegraphics[width=\hsize]{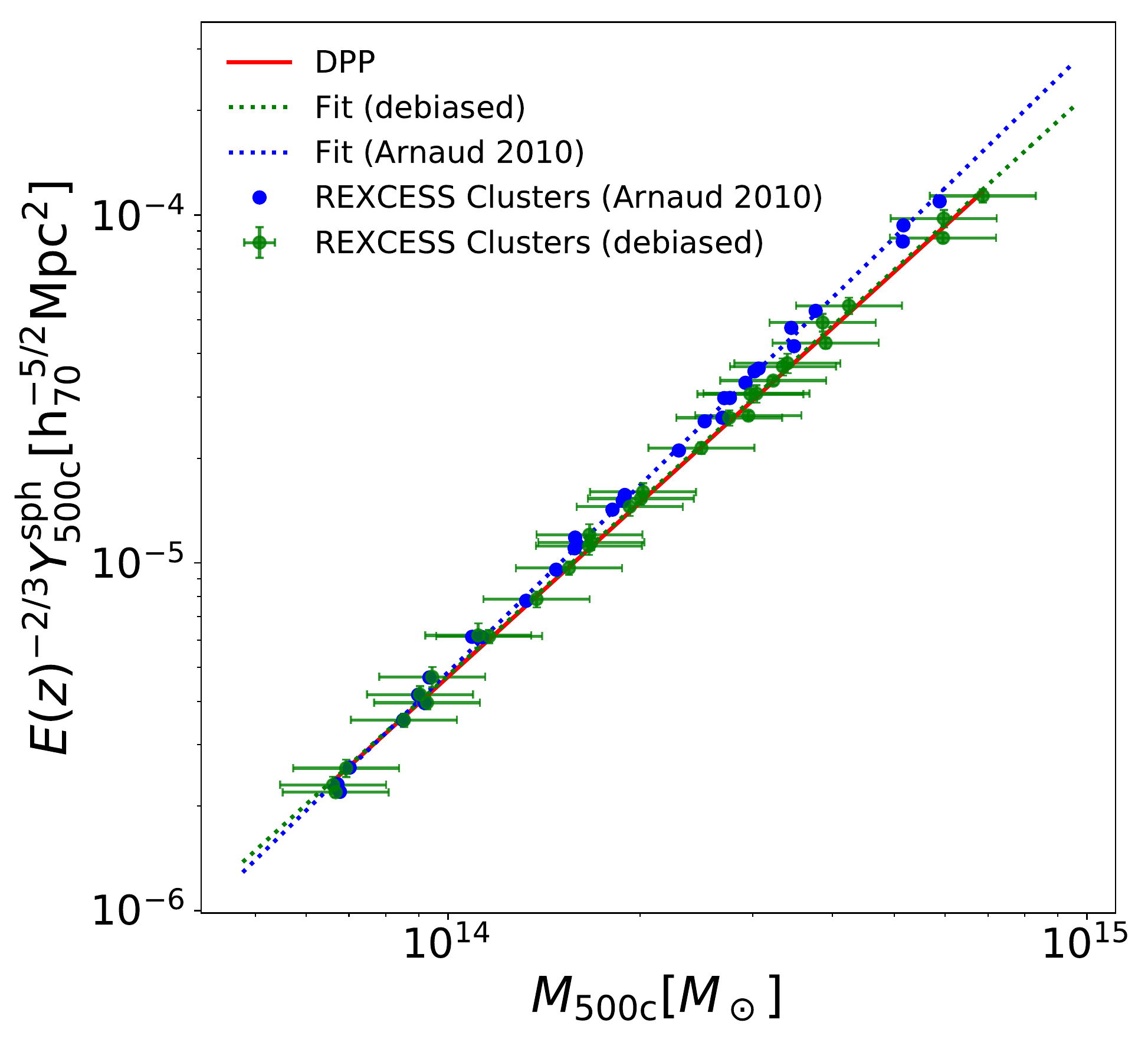}
\caption{The spherical volume-integrated Compton parameter, $Y_\text{sph},$ vs. mass, $M_{\rm 500c}$, for the REXCESS sample after correcting for hydrostatic bias (green dots) and the corresponding best-fit power-law relation (dashed green line). The analytical $Y_{\rm sph}(R_{\rm 500c})-M_{\rm 500c}$ relation derived  from the DPP model (solid red line) is also shown. The biased $Y_\text{sph}(R_\text{500c})$ and $M_\text{500c}$ (blue dots) and the corresponding best-fit power-law relation (dashed blue line) from \citet{2010A&A...517A..92A} are plotted for comparison.}\label{fig:Y-M relation}
\end{figure}

The spherical volume-integrated Compton parameter, $Y_\text{sph}$, of a cluster is the integral of the gas's thermal pressure profile over a spherical region and is defined as:
\begin{equation}\label{eq:Compton_y} 
Y_\text{sph}(R)=\frac{\sigma_\text{T}}{m_\text{e}c^2}\int_0^R4\pi P_\text{e}(r)r^2\mathrm{d}r,
\end{equation} 
where $\sigma_\text{T}$ is the Thomson cross-section, $m_\text{e}$ is the electron mass, and $P_\text{e}$ is the thermal electron pressure. Since the pressure is directly related to the depth of cluster gravitation potential, the integrated Compton parameter, $Y_\text{sph}$, is closely related to the mass of the cluster. Studies \citep[e.g.][]{doi:10.1111/j.1365-2966.2004.07463.x,0004-637X-650-2-538} find a low intrinsic scatter in the relation between integrated Compton parameter and cluster mass, indicating that the Compton parameter $Y_\text{sph}$ serves as a good proxy for cluster mass. The $Y_\text{sph}-M$ relation was previously modeled with a power-law \citep{2006ApJ...650..128K,2007ApJ...668....1N,2007A&A...474L..37A}. Accordingly, we parameterize the $Y_\text{sph}(R_{\rm 500c})-M_\text{500c}$ relation as 
\begin{equation}\label{eq:Y-M}
h(z)^{-2/3}Y_\text{sph}(R_\text{500c})=10^A\left[\frac{M_\text{500c}}{3\times10^{14}M_{\odot}}\right]^{\alpha}h_\text{70}^{-5/2}\text{Mpc}^2.
\end{equation}  

We fit Eq.~\ref{eq:Y-M} to the X-ray-measured Compton parameter and the biased X-ray hydrostatic masses of the REXCESS sample and find that $\alpha=1.790\pm0.015$, and $A=-4.739\pm0.003$. The $Y_\text{sph}-M$ relation can be derived from the UPP model by combining Eq.~\ref{eq:upp} for the UPP and Eq.~\ref{eq:Compton_y} and gives $\alpha=1.787$, and $A=-4.745$. The analytical calculations based on UPP and direct fits to observation data are in excellent agreement: both claimed a deviation from the slope predicted by self-similarity, $\alpha_\mathrm{s}=5/3$, of approximately $\Delta\alpha=\alpha-\alpha_\mathrm{s}\approx0.12$. Notice this deviation  $\Delta\alpha$ corresponds to the $\alpha_P(x)$ for the pressure model, which is characterized by a function of cluster mass and radius, however, \citet{2010A&A...517A..92A} showed this term can be approximated by a constant in the calculation of the spherical Compton signal and only causes a difference of $\leq$ 1\% for clusters in the mass range [$10^{14}M_\odot$, $10^{15}M_\odot$].

However, the hydrostatic masses used for constructing the UPP model are systematically underestimated, which means that the cluster radii are also biased. Integrating an X-ray-measured pressure profile over a biased volume leads to a biased Compton signal. We apply the rescaling methods discussed in Section \ref{subsec:HSBforP} to the GNFW fits to scaled pressure profiles and correct the X-ray measured radii of every REXCESS cluster, and correct the bias in the Compton parameter derived from X-ray measurements.
We also calculate $Y_\text{sph}$ analytically by integrating our \pressuremodel\ over the cluster within the radius $r=xR_\text{500c}$
\begin{align}\label{eq:25}
Y_\text{sph}(xR_\text{500c})&=
\frac{4\pi\sigma_\text{T}}{3m_\text{e}c^2}R_\text{500c}^3 P_\text{500c}\\ \notag
&\times\int_0^x3(x^\prime)^2\mathbb{P}(x^\prime)\left[\frac{M_\text{500c}}{3\times10^{14}h_{70}^{-1}M_{\odot}}\right]^{\alpha_\text{P}(x)}\mathrm dx^\prime,
\end{align}
then we simplify the integral, getting
\begin{equation}\label{eq:simplified Y-M}
h(z)^{-2/3}Y_\text{sph}(xR_\text{500c})=C(x)\left[\frac{M_\text{500c}}{3\times10^{14}h_\text{70}^{-1}M_{\odot}}\right]^{\alpha},
\end{equation}
where $\alpha=5/3$ given by $P_\text{500c}R_\text{500c}^3$, since $\alpha_{\text{P}(x)}$ is set to be 0 in \pressuremodel{} and has no contribution to $\alpha$, and
\begin{align}\label{eq:27}
C(x)&=2.925\times10^{-5}I(x)h_{70}^{-1}\text{Mpc}^2 \notag, \\
I(x)&=\int_0^x3\mathbb{P}(x^\prime)(x^\prime)^2\mathrm dx^\prime.
\end{align}
We use the value for the parameters $P_{0},c_{500},\alpha,\beta,\gamma$ of $P_\mathrm{m}$ reported in Table \ref{tab:pressure_params} to get $I(1)=0.554$. Rewriting $C(1)$ to the logarithmic form $10^A$, we get $A=-4.790$, along with $\alpha=5/3,$ as previously discussed for $Y_\mathrm{sph}-M$ relation, which agrees well with the direct fit to the REXCESS sample after correcting for hydrostatic bias: $\alpha=1.673\pm0.014$ and $A=-4.786\pm0.004$. 
 
In Figure \ref{fig:Y-M relation}, we plot fits for the integrated Compton signal versus cluster mass after correction for hydrostatic bias and analytical calculated $Y-M$ relation based on \pressuremodel. For comparison, we also plot $Y-M$ relation reported in \citet{2010A&A...517A..92A}. 

The corrected $ Y-M $ relation leads to smaller $ Y $ values at a given $ M $ compared to the UPP model, which indicates that our $ Y-M $ relation predicts a higher mass for the observed cluster given the same measured Compton signal. The new fit also shows a negligible difference from analytical results based on \pressuremodel. The value of the best-fit slope is close to the self-similar scaling with a tiny deviation $\Delta\alpha=1.673-5/3=0.006$.

We find no evidence for a power-law index of the $Y_{\rm sph}-M_{\rm 500c}$ relation which deviates from the predictions of self-similarity, which is consistent with a small deviation from standard self-similarity of the $Y_\text{sph}$-mass scaling relation in \cite{2017MNRAS.469.3069G}. The disappearance of the deviation from self-similarity is mainly due to the dependence of hydrostatic bias on cluster mass. We find changes in the spherical Compton signal of clusters in the REXCESS sample after adjusting for hydrostatic bias are much less significant, $< 2\%$ compared to the correction of cluster masses, which means the shift in cluster masses is the key factor for the modification on the power-law index of the $Y-M$ relation. Notice that the relation of $M_\text{500c}^\mathrm{True}$ vs. $M_\text{500c}^\mathrm{X-ray}$ for the REXCESS sample we derived yields $B_M\propto M^{\simeq0.08},$ equal to the shift of cluster mass after correction. To a great extent, this explains the variation -- around 0.12 -- of the power-law index of the $Y-M$ relation after adjusting for hydrostatic bias. 

The slope of the $Y-M$ relation being consistent with the standard self-similar model after adjusting for hydrostatic bias indicates that the studies that claim deviations from self-similarity in mass scaling relation like $L_\mathrm{X}-M$ and $Y_\mathrm{X}-M$ \citep[e.g.][]{2003MNRAS.342..287A, 2007A&A...474L..37A} may need to be revised as their results are also affected by similar X-ray mass biases. Additionally, the existence of self-similarity in the mass independence of our pressure model and the Y-M relation makes us more confident in extrapolating out the DPP model to higher redshifts in the calculation of thermal SZ power spectrum by assuming a self-similarity in the redshift dependence according to the standard self-similar model. We also note that the REXCESS sample has a limited mass range. Further confirmation of self-similarity in the $Y-M$ relation requires joint work of simulations and further observations with extended mass range.

\subsection{Thermal SZ Angular Power Spectrum}\label{subsec:tsz}

\begin{figure}
\includegraphics[width=\hsize]{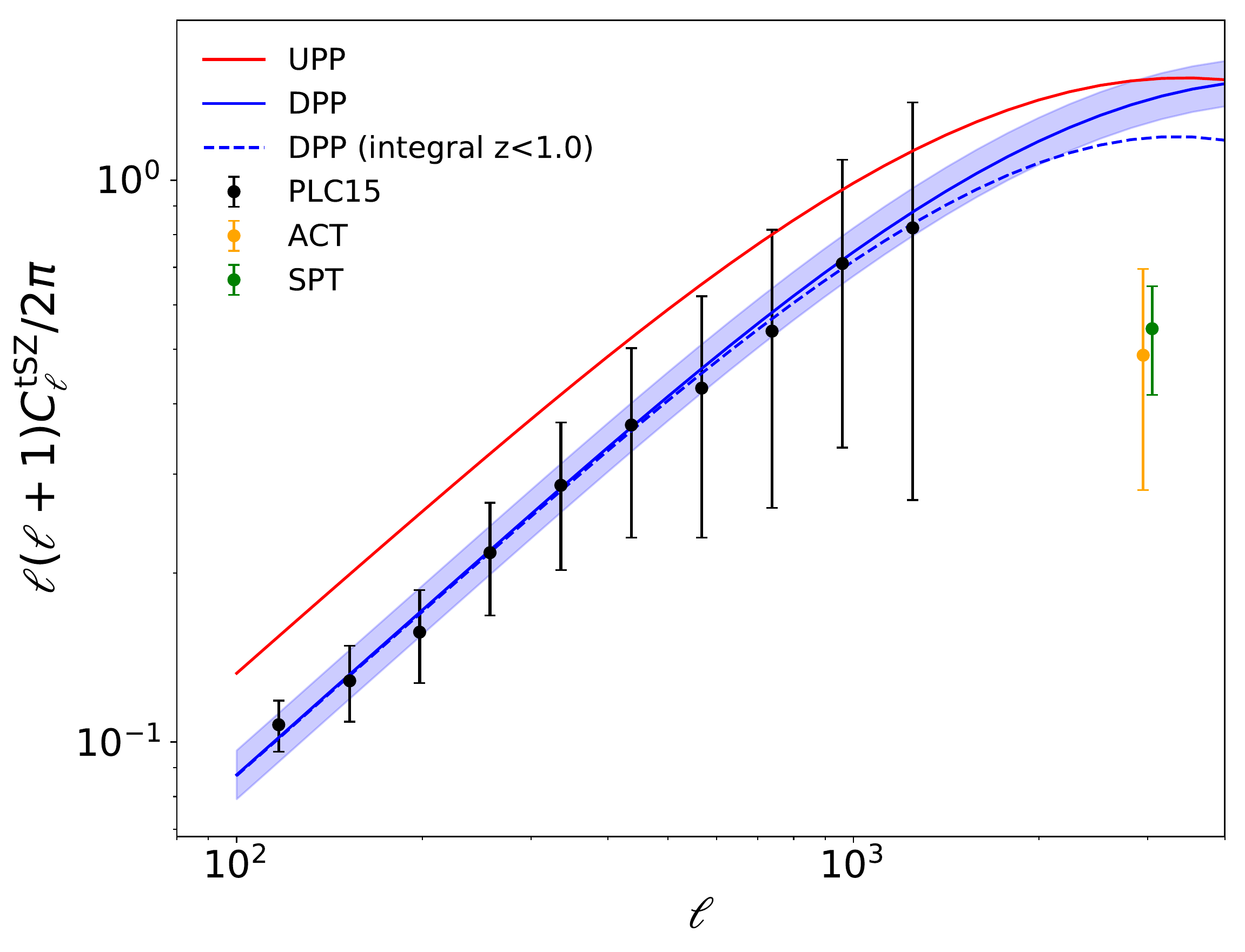}
\caption{Predictions for the one-halo term of the tSZ power spectrum calculated with the UPP model (red line) and the \pressuremodel\ model (blue line). The tSZ power spectrum calculated with Equation~\ref{eq:tSZ} integrated from $z=0.0$ to $z=1.0$ based on the \pressuremodel\ model is plotted for comparison (dashed blue line). Planck 2015 analysis of the tSZ power spectrum(black dots) with error bars due to uncertainties of foreground contamination and statistical errors. ACT (orange dot with error bar), and SPT (green dot with error bar) values correspond to $\mathscr{l}=3000$ are also shown, but they have been shifted in the plot for clarity. All tSZ data are rescaled to 146GHz for direct comparison, the uncertainty of the tSZ power spectrum (blue semitransparent band) is due to the uncertainty in the pressure profile used for the integral.}\label{fig:tSZ}
\end{figure}

The tSZ power spectrum is a powerful probe of cosmology  and can provide promising constraints on cosmological parameters: $C_\mathscr{l}\propto \sigma_\text{8}^{7-9}$ \citep[e.g.][]{2002MNRAS.336.1256K, 0004-637X-725-2-1452,0004-637X-727-2-94}. Since clusters are the dominant source of tSZ anisotropies, due to the number density of clusters and the gas thermal pressure profile, the tSZ power spectrum can be adequately modeled by an approach referred to as the halo formalism \citep[e.g.][]{1988MNRAS.233..637C, 1538-4357-526-1-L1}. 

The tSZ angular power spectrum at a multipole moment, $\mathscr{l}$, for the one-halo term is given by 
\begin{equation}\label{eq:tSZ}
C_\mathscr{l}^\mathrm{tSZ}=f^2(\nu)\int_z\frac{\mathrm dV}{\mathrm dz}\int_M\frac{\mathrm dn(M,z)}{\mathrm dM}|\tilde{y}_\mathscr{l}(M,z)|^2{\mathrm dMdz} ,
\end{equation}
where $f(\nu)=x\coth(x/2)-4$ is the spectral dependence with $x=h\nu/(k_\text{B}T_\text{CMB})$. Integration over redshift and mass are carried out from $z=0.0$ to $z=6.0$ and from $M=10^{10}M_{\odot}$ to $M=10^{16}M_{\odot}$ respectively. For the differential halo mass function $\mathrm dn(M,z)/\mathrm dM$, we adopt the fitting function from \citet{2008ApJ...688..709T} based on N-body simulations. 

Following \citet{2002MNRAS.336.1256K}, the 2D Fourier transform of the projected Compton $y$-parameter, $\tilde{y}_l(M,z)$ is given by 
\begin{equation}\label{eq:29}
\tilde{y}_\mathscr{l}(M,z)=\frac{4\pi r_\text{s}}{\mathscr{l}_\text{s}^2}\frac{\sigma_\text{T}}{m_\text{e}c^2}\int x^2P_\text{e}(x)\frac{\sin(\mathscr{l}x/\mathscr{l}_\text{s})}{\mathscr{l}x/\mathscr{l}_\text{s}}\mathrm dx,
\end{equation}
with the limber approximation \citep{1953ApJ...117..134L}, which is used to relate the angular correlation function to the corresponding three-dimensional spatial clustering in an approximate way and to avoid spherical Bessel function calculations, where $x=r/r_\text{s}$ is a scaled dimensionless radius, $r_\text{s}$ is characteristic radius for a NFW profile defined by $R_\text{500c}/c_\text{500c}$, and we use  average halo concentrations, $c_\text{500c},$ calibrated as a function of  cluster mass and redshift from \citet{2015ApJ...799..108D}. The corresponding angular wave number $\mathscr{l}_\text{s}=d_\text{A}/r_\text{s}$, where $d_\text{A}(z)$ is the proper angular-diameter distance at redshift $z$. $P_\text{e}(x)$ is the electron pressure we've discussed about Section~\ref{subsec:mass_relation}. The integral is carried out within a spherical region with radius $R\sim 4R_{500c}$.

In Figure~\ref{fig:tSZ}, we compare the measured tSZ power spectrum to the one-halo term predicted by the UPP and \pressuremodel{} models. Our predictions for the tSZ spectrum are made by assuming the fiducial parameters $\Omega_m=0.3,\ \Omega_\Lambda=0.7,\ \Omega_b=0.045,\ h=0.7,\ n_s=0.96,\ \sigma_8=0.8$. We use measurements of the tSZ power spectrum from the analysis of ACT \citep{2013JCAP...07..025D}, SPT \citep{0004-637X-799-2-177}, and Planck \citep{2016A&A...594A..22P}, all rescaled to 146 GHz, at which $f^2(\nu)=1$, for direct comparison. The uncertainties in the Planck 2015 data points account for statistical and systematic errors, as well as modeling uncertainties associated with correcting for foreground contamination. 

The tSZ power spectrum derived from the original UPP model predicts much higher values than observational data while our \pressuremodel\ model leads to the tSZ power spectrum matches the tSZ data of Planck within $1-\sigma$ uncertainty for multipoles $100\leq \mathscr{l}\leq 1300$. However, the tSZ power spectrum calculated with our \pressuremodel\ model still shows a significant tension with ACT and SPT data at $\mathscr{l}=3000$. Our work shows adjusting ICM pressure profiles for hydrostatic bias due to non-thermal pressure has a significant effect on lowering the amplitude of the power spectrum by  30-40\%. This is in agreement with other work studying the change in the shape of the tSZ power spectrum after including the effect of non-thermal pressure \citep[e.g.][]{0004-637X-725-2-1452,2010ApJ...725...91B, 0004-637X-727-2-94,0004-637X-758-2-75}.

In the analytical calculation of the tSZ power spectrum, we extrapolate our pressure model to redshifts as high as $z=6.0$, even though our pressure model is built on simulation data from a low redshift snapshot ($z=0.1$). However, we show in Figure \ref{fig:tSZ} that galaxy clusters of redshift $z\geq1.0$ will not significantly affect our calculation of the tSZ power spectrum at $\mathscr{l}\leq1300$. The tSZ power spectrum at $\mathscr{l}\geq3000$ shows it is more sensitive to the high redshift clusters, which may indicate that redshift dependence could potentially lower the tension between our calculation of the tSZ power spectrum and high-multipole observations.

\section{Discussion and conclusions}\label{sec:discussionandconclusions}

In this paper, we presented a simulation-based model to characterize the relation between the ``true" masses and the X-ray-estimated hydrostatic masses of galaxy clusters. We use X-ray masses measured from synthetic images of simulated clusters drawn from the IllustrisTNG, the BAHAMAS, and the MACSIS simulations \citep{2020arXiv200111508B} to fit a power-law relation for $M_{\rm 500c}^{\rm True}$ v.s. $M_{\rm 500c}^{\rm X-ray}$. We then use this model to correct the X-ray measured hydrostatic masses for the 31 clusters in the REXCESS sample:

\begin{enumerate}
\item We find that X-ray-measured hydrostatic masses underestimate masses of the clusters in the REXCESS sample by around $7\%$ on average and that the bias increases with mass from $\approx 0\%$ at $M_{500c}^{\rm X-ray}=10^{14}M_\odot$ to $\approx 15\%$ at $M_{500c}^{\rm X-ray}=10^{15}M_\odot,$ showing the same significant mass dependence as the simulation results. 
\item The significant scatter in simulation results has been incorporated into our model. This scatter also induces non-negligible uncertainties in the corrected of masses of individual REXCESS clusters, around $\pm20\%$.
\end{enumerate}

In this work, we assume mass bias does not or only weakly depends on the redshift. The REXCESS sample spans a redshift range of $0<z<0.2$ and our correction is based only on $z=0.1$ snapshots. To study the dependence of mass bias on redshift, one could look into more snapshots of the simulations we used. As we mention in Section~\ref{subsec:mupp}, the dynamical states of different of RECXESS clusters could vary significantly, which could also be considered a selection criterion in addition to the cluster mass. Furthering modeling may be improved by accounting for dynamical state when correcting X-ray-measured hydrostatic masses.  

We discussed how the mass bias we found transfers to other X-ray observables. Scaling relations between cluster mass, radius, and characteristic pressure  ($R\propto M^{1/3}$, $P\propto M^{2/3}$), enable a convenient correction of GNFW fits to scaled pressure profiles, through the modification of the $P_0$ and $c_{500}$ parameters.

We adjusted the universal galaxy cluster pressure profile for hydrostatic mass bias through recalibrating the scaled pressure profiles of each cluster in the REXCESS samples used to construct the UPP model:
\begin{enumerate}
\item In our updated pressure model, \pressuremodel, pressures are 5\% lower than the UPP model in the inner region of the clusters, and 15$\%$ lower at $R_\text{500c}$. 

\item We achieve a good agreement on a small value of $\alpha_\text{P}$ in the respective study of pressure model and $Y_\text{sph}-M$ relation, which implies standard self-similarity still stands for the scaling relation of the adjusted universal pressure model and the $Y_\text{sph}-M$ relation.

\item An analytical calculation of the thermal SZ angular power spectrum derived from \pressuremodel\ is consistent with the analysis of Planck thermal SZ survey data without requiring extreme cosmological parameters.
\end{enumerate}

Many avenues remain for future work on this topic. Our analysis is restricted to late times, meaning that we do not explore the redshift dependence of hydrostatic mass bias. Analysis that incorporates redshift evolution would likely lead to more accurate cosmological constraints from the tSZ power spectrum. Similar to the UPP, our \pressuremodel{} does not differentiate between relaxed and unrelaxed clusters or cool core and non-cool-core clusters. The impact of hydrostatic mass bias on these clusters sub-categories has not yet been determined. Lastly, we note that even our corrected fit cannot simultaneously match the $\mathscr{l}=3000$ tSZ power spectrum measurements from ACT and SPT. This discrepancy remains an open question.

\section*{Acknowledgements}

We thank Arya Farahi, Matthew Hasselfield, Matt Hilton, Kaylea Nelson, and Andrey Kravtsov for useful discussions. We thank David Barnes for the mass data of simulated clusters drawn from the IllustrisTNG, BAHAMAS, and MACSIS simulations with Mock-X analysis framework. PM was supported by the Kavli Institute for Cosmological Physics at the University of Chicago through grant PHY-1125897, and AST-1714658, and an endowment from the Kavli Foundation and its founder, Fred Kavli.

\bibliographystyle{aasjournal}
\bibliography{ref}

\begin{thebibliography}{}
\expandafter\ifx\csname natexlab\endcsname\relax\def\natexlab#1{#1}\fi
\providecommand{\url}[1]{\href{#1}{#1}}

\bibitem[{{Allen} {et~al.}(2011){Allen}, {Evrard}, \&
  {Mantz}}]{2011ARA&A..49..409A}
{Allen}, S.~W., {Evrard}, A.~E., \& {Mantz}, A.~B. 2011, \araa, 49, 409

\bibitem[{{Allen} {et~al.}(2003){Allen}, {Schmidt}, {Fabian}, \&
  {Ebeling}}]{2003MNRAS.342..287A}
{Allen}, S.~W., {Schmidt}, R.~W., {Fabian}, A.~C., \& {Ebeling}, H. 2003,
  \mnras, 342, 287

\bibitem[{{Arnaud} {et~al.}(2007){Arnaud}, {Pointecouteau}, \&
  {Pratt}}]{2007A&A...474L..37A}
{Arnaud}, M., {Pointecouteau}, E., \& {Pratt}, G.~W. 2007, \aap, 474, L37

\bibitem[{{Arnaud} {et~al.}(2010){Arnaud}, {Pratt}, {Piffaretti},
  {B{\"o}hringer}, {Croston}, \& {Pointecouteau}}]{2010A&A...517A..92A}
{Arnaud}, M., {Pratt}, G.~W., {Piffaretti}, R., {et~al.} 2010, \aap, 517, A92

\bibitem[{{Barnes} {et~al.}(2017){Barnes}, {Kay}, {Henson}, {McCarthy},
  {Schaye}, \& {Jenkins}}]{2017MNRAS.465..213B}
{Barnes}, D.~J., {Kay}, S.~T., {Henson}, M.~A., {et~al.} 2017, \mnras, 465, 213

\bibitem[{{Barnes} {et~al.}(2020){Barnes}, {Vogelsberger}, {Pearce}, {Pop},
  {Kannan}, {Cao}, {Kay}, \& {Hernquist}}]{2020arXiv200111508B}
{Barnes}, D.~J., {Vogelsberger}, M., {Pearce}, F.~A., {et~al.} 2020, arXiv
  e-prints, arXiv:2001.11508

\bibitem[{{Battaglia} {et~al.}(2012){Battaglia}, {Bond}, {Pfrommer}, \&
  {Sievers}}]{2012ApJ...758...74B}
{Battaglia}, N., {Bond}, J.~R., {Pfrommer}, C., \& {Sievers}, J.~L. 2012, \apj,
  758, 74

\bibitem[{Battaglia {et~al.}(2012)Battaglia, Bond, Pfrommer, \&
  Sievers}]{0004-637X-758-2-75}
Battaglia, N., Bond, J.~R., Pfrommer, C., \& Sievers, J.~L. 2012, The
  Astrophysical Journal, 758, 75.
\newblock \url{http://stacks.iop.org/0004-637X/758/i=2/a=75}

\bibitem[{{Battaglia} {et~al.}(2010){Battaglia}, {Bond}, {Pfrommer}, {Sievers},
  \& {Sijacki}}]{2010ApJ...725...91B}
{Battaglia}, N., {Bond}, J.~R., {Pfrommer}, C., {Sievers}, J.~L., \& {Sijacki},
  D. 2010, \apj, 725, 91

\bibitem[{Battaglia {et~al.}(2016)Battaglia, Leauthaud, Miyatake, Hasselfield,
  Gralla, Allison, Bond, Calabrese, Crichton, Devlin, Dunkley, D{\"u}nner,
  Erben, Ferrara, Halpern, Hilton, Hill, Hincks, Hlo{\v{z}}ek, Huffenberger,
  Hughes, Kneib, Kosowsky, Makler, Marriage, Menanteau, Miller, Moodley,
  Moraes, Niemack, Page, Shan, Sehgal, Sherwin, Sievers, Sif{\'{o}}n, Spergel,
  Staggs, Taylor, Thornton, van Waerbeke, \& Wollack}]{Battaglia_2016}
Battaglia, N., Leauthaud, A., Miyatake, H., {et~al.} 2016, Journal of Cosmology
  and Astroparticle Physics, 2016, 013.
\newblock \url{https://doi.org/10.1088%2F1475-7516%2F2016%2F08%2F013}

\bibitem[{{Bautz} {et~al.}(2009){Bautz}, {Miller}, {Sanders}, {Arnaud},
  {Mushotzky}, {Porter}, {Hayashida}, {Henry}, {Hughes}, {Kawaharada},
  {Makashima}, {Sato}, \& {Tamura}}]{2009PASJ...61.1117B}
{Bautz}, M.~W., {Miller}, E.~D., {Sanders}, J.~S., {et~al.} 2009, \pasj, 61,
  1117

\bibitem[{{B{\"o}hringer} {et~al.}(2007){B{\"o}hringer}, {Schuecker}, {Pratt},
  {Arnaud}, {Ponman}, {Croston}, {Borgani}, {Bower}, {Briel}, {Collins},
  {Donahue}, {Forman}, {Finoguenov}, {Geller}, {Guzzo}, {Henry}, {Kneissl},
  {Mohr}, {Matsushita}, {Mullis}, {Ohashi}, {Pedersen}, {Pierini}, {Quintana},
  {Raychaudhury}, {Reiprich}, {Romer}, {Rosati}, {Sabirli}, {Temple}, {Viana},
  {Vikhlinin}, {Voit}, \& {Zhang}}]{2007A&A...469..363B}
{B{\"o}hringer}, H., {Schuecker}, P., {Pratt}, G.~W., {et~al.} 2007, \aap, 469,
  363

\bibitem[{{Br{\"u}ggen} \& {Vazza}(2015)}]{2015ASSL..407..599B}
{Br{\"u}ggen}, M., \& {Vazza}, F. 2015, Astrophysics and Space Science Library,
  Vol. 407, {Turbulence in the Intracluster Medium}, ed. A.~{Lazarian}, E.~M.
  {de Gouveia Dal Pino}, \& C.~{Melioli}, 599

\bibitem[{{Churazov} {et~al.}(2008){Churazov}, {Forman}, {Vikhlinin},
  {Tremaine}, {Gerhard}, \& {Jones}}]{2008MNRAS.388.1062C}
{Churazov}, E., {Forman}, W., {Vikhlinin}, A., {et~al.} 2008, \mnras, 388, 1062

\bibitem[{{Cole} \& {Kaiser}(1988)}]{1988MNRAS.233..637C}
{Cole}, S., \& {Kaiser}, N. 1988, \mnras, 233, 637

\bibitem[{{Croston} {et~al.}(2008){Croston}, {Pratt}, {B{\"o}hringer},
  {Arnaud}, {Pointecouteau}, {Ponman}, {Sand erson}, {Temple}, {Bower}, \&
  {Donahue}}]{2008A&A...487..431C}
{Croston}, J.~H., {Pratt}, G.~W., {B{\"o}hringer}, H., {et~al.} 2008, \aap,
  487, 431

\bibitem[{Da~Silva {et~al.}(2004)Da~Silva, Kay, Liddle, \&
  Thomas}]{doi:10.1111/j.1365-2966.2004.07463.x}
Da~Silva, A.~C., Kay, S.~T., Liddle, A.~R., \& Thomas, P.~A. 2004, Monthly
  Notices of the Royal Astronomical Society, 348, 1401.
\newblock \url{+ http://dx.doi.org/10.1111/j.1365-2966.2004.07463.x}

\bibitem[{{Diemer} \& {Kravtsov}(2015)}]{2015ApJ...799..108D}
{Diemer}, B., \& {Kravtsov}, A.~V. 2015, \apj, 799, 108

\bibitem[{{Dolag} {et~al.}(2009){Dolag}, {Borgani}, {Murante}, \&
  {Springel}}]{2009MNRAS.399..497D}
{Dolag}, K., {Borgani}, S., {Murante}, G., \& {Springel}, V. 2009, \mnras, 399,
  497

\bibitem[{{Dunkley} {et~al.}(2013){Dunkley}, {Calabrese}, {Sievers}, {Addison},
  {Battaglia}, {Battistelli}, {Bond}, {Das}, {Devlin}, {D{\"u}nner}, {Fowler},
  {Gralla}, {Hajian}, {Halpern}, {Hasselfield}, {Hincks}, {Hlozek}, {Hughes},
  {Irwin}, {Kosowsky}, {Louis}, {Marriage}, {Marsden}, {Menanteau}, {Moodley},
  {Niemack}, {Nolta}, {Page}, {Partridge}, {Sehgal}, {Spergel}, {Staggs},
  {Switzer}, {Trac}, \& {Wollack}}]{2013JCAP...07..025D}
{Dunkley}, J., {Calabrese}, E., {Sievers}, J., {et~al.} 2013, \jcap, 7, 025

\bibitem[{{Ensslin} {et~al.}(1997){Ensslin}, {Biermann}, {Kronberg}, \&
  {Wu}}]{1997ApJ...477..560E}
{Ensslin}, T.~A., {Biermann}, P.~L., {Kronberg}, P.~P., \& {Wu}, X.-P. 1997,
  \apj, 477, 560

\bibitem[{{Evrard}(1990)}]{1990ApJ...363..349E}
{Evrard}, A.~E. 1990, \apj, 363, 349

\bibitem[{{Foreman-Mackey} {et~al.}(2013){Foreman-Mackey}, {Hogg}, {Lang}, \&
  {Goodman}}]{2013PASP..125..306F}
{Foreman-Mackey}, D., {Hogg}, D.~W., {Lang}, D., \& {Goodman}, J. 2013, \pasp,
  125, 306

\bibitem[{{Foreman-Mackey} {et~al.}(2019){Foreman-Mackey}, {Farr}, {Sinha},
  {Archibald}, {Hogg}, {Sanders}, {Zuntz}, {Williams}, {Nelson}, {de
  Val-Borro}, {Erhardt}, {Pashchenko}, \& {Pla}}]{2019JOSS....4.1864F}
{Foreman-Mackey}, D., {Farr}, W., {Sinha}, M., {et~al.} 2019, The Journal of
  Open Source Software, 4, 1864

\bibitem[{George {et~al.}(2015)George, Reichardt, Aird, Benson, Bleem,
  Carlstrom, Chang, Cho, Crawford, Crites, de~Haan, Dobbs, Dudley, Halverson,
  Harrington, Holder, Holzapfel, Hou, Hrubes, Keisler, Knox, Lee, Leitch,
  Lueker, Luong-Van, McMahon, Mehl, Meyer, Millea, Mocanu, Mohr, Montroy,
  Padin, Plagge, Pryke, Ruhl, Schaffer, Shaw, Shirokoff, Spieler, Staniszewski,
  Stark, Story, van Engelen, Vanderlinde, Vieira, Williamson, \&
  Zahn}]{0004-637X-799-2-177}
George, E.~M., Reichardt, C.~L., Aird, K.~A., {et~al.} 2015, The Astrophysical
  Journal, 799, 177.
\newblock \url{http://stacks.iop.org/0004-637X/799/i=2/a=177}

\bibitem[{{George} {et~al.}(2009){George}, {Fabian}, {Sanders}, {Young}, \&
  {Russell}}]{2009MNRAS.395..657G}
{George}, M.~R., {Fabian}, A.~C., {Sanders}, J.~S., {Young}, A.~J., \&
  {Russell}, H.~R. 2009, \mnras, 395, 657

\bibitem[{{Gupta} {et~al.}(2017){Gupta}, {Saro}, {Mohr}, {Dolag}, \&
  {Liu}}]{2017MNRAS.469.3069G}
{Gupta}, N., {Saro}, A., {Mohr}, J.~J., {Dolag}, K., \& {Liu}, J. 2017, \mnras,
  469, 3069

\bibitem[{{Hasselfield} {et~al.}(2013){Hasselfield}, {Hilton}, {Marriage},
  {Addison}, {Barrientos}, {Battaglia}, {Battistelli}, {Bond}, {Crichton},
  {Das}, {Devlin}, {Dicker}, {Dunkley}, {D{\"u}nner}, {Fowler}, {Gralla},
  {Hajian}, {Halpern}, {Hincks}, {Hlozek}, {Hughes}, {Infante}, {Irwin},
  {Kosowsky}, {Marsden}, {Menanteau}, {Moodley}, {Niemack}, {Nolta}, {Page},
  {Partridge}, {Reese}, {Schmitt}, {Sehgal}, {Sherwin}, {Sievers}, {Sif{\'o}n},
  {Spergel}, {Staggs}, {Swetz}, {Switzer}, {Thornton}, {Trac}, \&
  {Wollack}}]{2013JCAP...07..008H}
{Hasselfield}, M., {Hilton}, M., {Marriage}, T.~A., {et~al.} 2013, \jcap, 7,
  008

\bibitem[{Henson {et~al.}(2017)Henson, Barnes, Kay, McCarthy, \&
  Schaye}]{doi:10.1093/mnras/stw2899}
Henson, M.~A., Barnes, D.~J., Kay, S.~T., McCarthy, I.~G., \& Schaye, J. 2017,
  Monthly Notices of the Royal Astronomical Society, 465, 3361.
\newblock \url{+ http://dx.doi.org/10.1093/mnras/stw2899}

\bibitem[{{Hilton} {et~al.}(2018){Hilton}, {Hasselfield}, {Sif{\'o}n},
  {Battaglia}, {Aiola}, {Bharadwaj}, {Bond}, {Choi}, {Crichton}, {Datta},
  {Devlin}, {Dunkley}, {D{\"u}nner}, {Gallardo}, {Gralla}, {Hincks}, {Ho},
  {Hubmayr}, {Huffenberger}, {Hughes}, {Koopman}, {Kosowsky}, {Louis},
  {Madhavacheril}, {Marriage}, {Maurin}, {McMahon}, {Miyatake}, {Moodley},
  {N{\ae}ss}, {Nati}, {Newburgh}, {Niemack}, {Oguri}, {Page}, {Partridge},
  {Schmitt}, {Sievers}, {Spergel}, {Staggs}, {Trac}, {van Engelen},
  {Vavagiakis}, \& {Wollack}}]{2018ApJS..235...20H}
{Hilton}, M., {Hasselfield}, M., {Sif{\'o}n}, C., {et~al.} 2018, \apjs, 235, 20

\bibitem[{{Hitomi Collaboration} {et~al.}(2018){Hitomi Collaboration},
  {Aharonian}, {Akamatsu}, {Akimoto}, {Allen}, {Angelini}, {Audard}, {Awaki},
  {Axelsson}, {Bamba}, {Bautz}, {Blandford}, {Brenneman}, {Brown}, {Bulbul},
  {Cackett}, {Canning}, {Chernyakova}, {Chiao}, {Coppi}, {Costantini}, {de
  Plaa}, {de Vries}, {den Herder}, {Done}, {Dotani}, {Ebisawa}, {Eckart},
  {Enoto}, {Ezoe}, {Fabian}, {Ferrigno}, {Foster}, {Fujimoto}, {Fukazawa},
  {Furuzawa}, {Galeazzi}, {Gallo}, {Gandhi}, {Giustini}, {Goldwurm}, {Gu},
  {Guainazzi}, {Haba}, {Hagino}, {Hamaguchi}, {Harrus}, {Hatsukade}, {Hayashi},
  {Hayashi}, {Hayashi}, {Hayashida}, {Hiraga}, {Hornschemeier}, {Hoshino},
  {Hughes}, {Ichinohe}, {Iizuka}, {Inoue}, {Inoue}, {Inoue}, {Ishida},
  {Ishikawa}, {Ishisaki}, {Iwai}, {Kaastra}, {Kallman}, {Kamae}, {Kataoka},
  {Katsuda}, {Kawai}, {Kelley}, {Kilbourne}, {Kitaguchi}, {Kitamoto},
  {Kitayama}, {Kohmura}, {Kokubun}, {Koyama}, {Koyama}, {Kretschmar}, {Krimm},
  {Kubota}, {Kunieda}, {Laurent}, {Lee}, {Leutenegger}, {Limousin},
  {Loewenstein}, {Long}, {Lumb}, {Madejski}, {Maeda}, {Maier}, {Makishima},
  {Markevitch}, {Matsumoto}, {Matsushita}, {McCammon}, {McNamara}, {Mehdipour},
  {Miller}, {Miller}, {Mineshige}, {Mitsuda}, {Mitsuishi}, {Miyazawa},
  {Mizuno}, {Mori}, {Mori}, {Mukai}, {Murakami}, {Mushotzky}, {Nakagawa},
  {Nakajima}, {Nakamori}, {Nakashima}, {Nakazawa}, {Nobukawa}, {Nobukawa},
  {Noda}, {Odaka}, {Ohashi}, {Ohno}, {Okajima}, {Ota}, {Ozaki}, {Paerels},
  {Paltani}, {Petre}, {Pinto}, {Porter}, {Pottschmidt}, {Reynolds},
  {Safi-Harb}, {Saito}, {Sakai}, {Sasaki}, {Sato}, {Sato}, {Sato}, {Sawada},
  {Schartel}, {Serlemtsos}, {Seta}, {Shidatsu}, {Simionescu}, {Smith}, {Soong},
  {Stawarz}, {Sugawara}, {Sugita}, {Szymkowiak}, {Tajima}, {Takahashi},
  {Takahashi}, {Takeda}, {Takei}, {Tamagawa}, {Tamura}, {Tanaka}, {Tanaka},
  {Tanaka}, {Tanaka}, {Tashiro}, {Tawara}, {Terada}, {Terashima}, {Tombesi},
  {Tomida}, {Tsuboi}, {Tsujimoto}, {Tsunemi}, {Tsuru}, {Uchida}, {Uchiyama},
  {Uchiyama}, {Ueda}, {Ueda}, {Uno}, {Urry}, {Ursino}, {Wang}, {Watanabe},
  {Werner}, {Wilkins}, {Williams}, {Yamada}, {Yamaguchi}, {Yamaoka},
  {Yamasaki}, {Yamauchi}, {Yamauchi}, {Yaqoob}, {Yatsu}, {Yonetoku},
  {Zhuravleva}, \& {Zoghbi}}]{2018PASJ...70....9H}
{Hitomi Collaboration}, {Aharonian}, F., {Akamatsu}, H., {et~al.} 2018, \pasj,
  70, 9

\bibitem[{{Hoekstra} {et~al.}(2015){Hoekstra}, {Herbonnet}, {Muzzin}, {Babul},
  {Mahdavi}, {Viola}, \& {Cacciato}}]{2015MNRAS.449..685H}
{Hoekstra}, H., {Herbonnet}, R., {Muzzin}, A., {et~al.} 2015, \mnras, 449, 685

\bibitem[{{Hoshino} {et~al.}(2010){Hoshino}, {Henry}, {Sato}, {Akamatsu},
  {Yokota}, {Sasaki}, {Ishisaki}, {Ohashi}, {Bautz}, {Fukazawa}, {Kawano},
  {Furuzawa}, {Hayashida}, {Tawa}, {Hughes}, {Kokubun}, \&
  {Tamura}}]{2010PASJ...62..371H}
{Hoshino}, A., {Henry}, J.~P., {Sato}, K., {et~al.} 2010, \pasj, 62, 371

\bibitem[{{Kawaharada} {et~al.}(2010){Kawaharada}, {Okabe}, {Umetsu},
  {Takizawa}, {Matsushita}, {Fukazawa}, {Hamana}, {Miyazaki}, {Nakazawa}, \&
  {Ohashi}}]{2010ApJ...714..423K}
{Kawaharada}, M., {Okabe}, N., {Umetsu}, K., {et~al.} 2010, \apj, 714, 423

\bibitem[{Kay {et~al.}(2012)Kay, Peel, Short, Thomas, Young, Battye, Liddle, \&
  Pearce}]{doi:10.1111/j.1365-2966.2012.20623.x}
Kay, S.~T., Peel, M.~W., Short, C.~J., {et~al.} 2012, Monthly Notices of the
  Royal Astronomical Society, 422, 1999.
\newblock \url{+ http://dx.doi.org/10.1111/j.1365-2966.2012.20623.x}

\bibitem[{Komatsu \& Kitayama(1999)}]{1538-4357-526-1-L1}
Komatsu, E., \& Kitayama, T. 1999, The Astrophysical Journal Letters, 526, L1.
\newblock \url{http://stacks.iop.org/1538-4357/526/i=1/a=L1}

\bibitem[{{Komatsu} \& {Seljak}(2002)}]{2002MNRAS.336.1256K}
{Komatsu}, E., \& {Seljak}, U. 2002, \mnras, 336, 1256

\bibitem[{{Kravtsov} {et~al.}(2006){Kravtsov}, {Vikhlinin}, \&
  {Nagai}}]{2006ApJ...650..128K}
{Kravtsov}, A.~V., {Vikhlinin}, A., \& {Nagai}, D. 2006, \apj, 650, 128

\bibitem[{Lau {et~al.}(2009)Lau, Kravtsov, \& Nagai}]{0004-637X-705-2-1129}
Lau, E.~T., Kravtsov, A.~V., \& Nagai, D. 2009, The Astrophysical Journal, 705,
  1129.
\newblock \url{http://stacks.iop.org/0004-637X/705/i=2/a=1129}

\bibitem[{{Lau} {et~al.}(2013){Lau}, {Nagai}, \&
  {Nelson}}]{2013ApJ...777..151L}
{Lau}, E.~T., {Nagai}, D., \& {Nelson}, K. 2013, \apj, 777, 151

\bibitem[{{Le Brun} {et~al.}(2014){Le Brun}, {McCarthy}, {Schaye}, \&
  {Ponman}}]{2014MNRAS.441.1270L}
{Le Brun}, A.~M.~C., {McCarthy}, I.~G., {Schaye}, J., \& {Ponman}, T.~J. 2014,
  \mnras, 441, 1270

\bibitem[{{Limber}(1953)}]{1953ApJ...117..134L}
{Limber}, D.~N. 1953, \apj, 117, 134

\bibitem[{{Marinacci} {et~al.}(2018){Marinacci}, {Vogelsberger}, {Pakmor},
  {Torrey}, {Springel}, {Hernquist}, {Nelson}, {Weinberger}, {Pillepich},
  {Naiman}, \& {Genel}}]{Marinacci_et_al_2018}
{Marinacci}, F., {Vogelsberger}, M., {Pakmor}, R., {et~al.} 2018, \mnras, 480,
  5113

\bibitem[{{McCarthy} {et~al.}(2017){McCarthy}, {Schaye}, {Bird}, \& {Le
  Brun}}]{McCarthy_et_al_2017}
{McCarthy}, I.~G., {Schaye}, J., {Bird}, S., \& {Le Brun}, A. M.~C. 2017,
  \mnras, 465, 2936

\bibitem[{{Medezinski} {et~al.}(2018){Medezinski}, {Battaglia}, {Umetsu},
  {Oguri}, {Miyatake}, {Nishizawa}, {Sif{\'o}n}, {Spergel}, {Chiu}, {Lin},
  {Bahcall}, \& {Komiyama}}]{2018PASJ...70S..28M}
{Medezinski}, E., {Battaglia}, N., {Umetsu}, K., {et~al.} 2018, \pasj, 70, S28

\bibitem[{{Miyatake} {et~al.}(2019){Miyatake}, {Battaglia}, {Hilton},
  {Medezinski}, {Nishizawa}, {More}, {Aiola}, {Bahcall}, {Bond}, {Calabrese},
  {Choi}, {Devlin}, {Dunkley}, {Dunner}, {Fuzia}, {Gallardo}, {Gralla},
  {Hasselfield}, {Halpern}, {Hikage}, {Hill}, {Hincks}, {Hlo{\v{z}}ek},
  {Huffenberger}, {Hughes}, {Koopman}, {Kosowsky}, {Louis}, {Madhavacheril},
  {McMahon}, {Mandelbaum}, {Marriage}, {Maurin}, {Miyazaki}, {Moodley},
  {Murata}, {Naess}, {Newburgh}, {Niemack}, {Nishimichi}, {Okabe}, {Oguri},
  {Osato}, {Page}, {Partridge}, {Robertson}, {Sehgal}, {Sherwin}, {Shirasaki},
  {Sievers}, {Sif{\'o}n}, {Simon}, {Spergel}, {Staggs}, {Stein}, {Takada},
  {Trac}, {Umetsu}, {van Engelen}, \& {Wollack}}]{2019ApJ...875...63M}
{Miyatake}, H., {Battaglia}, N., {Hilton}, M., {et~al.} 2019, \apj, 875, 63

\bibitem[{Nagai(2006)}]{0004-637X-650-2-538}
Nagai, D. 2006, The Astrophysical Journal, 650, 538.
\newblock \url{http://stacks.iop.org/0004-637X/650/i=2/a=538}

\bibitem[{{Nagai} {et~al.}(2007{\natexlab{a}}){Nagai}, {Kravtsov}, \&
  {Vikhlinin}}]{2007ApJ...668....1N}
{Nagai}, D., {Kravtsov}, A.~V., \& {Vikhlinin}, A. 2007{\natexlab{a}}, \apj,
  668, 1

\bibitem[{{Nagai} {et~al.}(2007{\natexlab{b}}){Nagai}, {Vikhlinin}, \&
  {Kravtsov}}]{2007ApJ...655...98N}
{Nagai}, D., {Vikhlinin}, A., \& {Kravtsov}, A.~V. 2007{\natexlab{b}}, \apj,
  655, 98

\bibitem[{{Naiman} {et~al.}(2018){Naiman}, {Pillepich}, {Springel},
  {Ramirez-Ruiz}, {Torrey}, {Vogelsberger}, {Pakmor}, {Nelson}, {Marinacci},
  {Hernquist}, {Weinberger}, \& {Genel}}]{Naiman_et_al_2018}
{Naiman}, J.~P., {Pillepich}, A., {Springel}, V., {et~al.} 2018, \mnras, 477,
  1206

\bibitem[{{Nelson} {et~al.}(2018){Nelson}, {Pillepich}, {Springel},
  {Weinberger}, {Hernquist}, {Pakmor}, {Genel}, {Torrey}, {Vogelsberger},
  {Kauffmann}, {Marinacci}, \& {Naiman}}]{Nelson_et_al_2018}
{Nelson}, D., {Pillepich}, A., {Springel}, V., {et~al.} 2018, \mnras, 475, 624

\bibitem[{Nelson {et~al.}(2014)Nelson, Lau, \& Nagai}]{0004-637X-792-1-25}
Nelson, K., Lau, E.~T., \& Nagai, D. 2014, The Astrophysical Journal, 792, 25.
\newblock \url{http://stacks.iop.org/0004-637X/792/i=1/a=25}

\bibitem[{{Nelson} {et~al.}(2012){Nelson}, {Rudd}, {Shaw}, \&
  {Nagai}}]{2012ApJ...751..121N}
{Nelson}, K., {Rudd}, D.~H., {Shaw}, L., \& {Nagai}, D. 2012, \apj, 751, 121

\bibitem[{{Penna-Lima} {et~al.}(2017){Penna-Lima}, {Bartlett}, {Rozo}, {Melin},
  {Merten}, {Evrard}, {Postman}, \& {Rykoff}}]{2017A&A...604A..89P}
{Penna-Lima}, M., {Bartlett}, J.~G., {Rozo}, E., {et~al.} 2017, \aap, 604, A89

\bibitem[{Pfanzagl \& Sheynin(1996)}]{10.1093/biomet/83.4.891}
Pfanzagl, J., \& Sheynin, O. 1996, Biometrika, 83, 891.
\newblock \url{https://doi.org/10.1093/biomet/83.4.891}

\bibitem[{{Pillepich} {et~al.}(2018){Pillepich}, {Nelson}, {Hernquist},
  {Springel}, {Pakmor}, {Torrey}, {Weinberger}, {Genel}, {Naiman}, {Marinacci},
  \& {Vogelsberger}}]{Phillepich_et_al_2018}
{Pillepich}, A., {Nelson}, D., {Hernquist}, L., {et~al.} 2018, \mnras, 475, 648

\bibitem[{{Planck Collaboration} {et~al.}(2016{\natexlab{a}}){Planck
  Collaboration}, {Ade}, {Aghanim}, {Arnaud}, {Ashdown}, {Aumont},
  {Baccigalupi}, {Banday}, {Barreiro}, {Barrena}, \&
  et~al.}]{2016A&A...594A..27P}
{Planck Collaboration}, {Ade}, P.~A.~R., {Aghanim}, N., {et~al.}
  2016{\natexlab{a}}, \aap, 594, A27

\bibitem[{{Planck Collaboration} {et~al.}(2016{\natexlab{b}}){Planck
  Collaboration}, {Aghanim}, {Arnaud}, {Ashdown}, {Aumont}, {Baccigalupi},
  {Banday}, {Barreiro}, {Bartlett}, {Bartolo}, \& et~al.}]{2016A&A...594A..22P}
{Planck Collaboration}, {Aghanim}, N., {Arnaud}, M., {et~al.}
  2016{\natexlab{b}}, \aap, 594, A22

\bibitem[{{Rasia} {et~al.}(2004){Rasia}, {Tormen}, \&
  {Moscardini}}]{2004MNRAS.351..237R}
{Rasia}, E., {Tormen}, G., \& {Moscardini}, L. 2004, \mnras, 351, 237

\bibitem[{Rasia {et~al.}(2012)Rasia, Meneghetti, Martino, Borgani, Bonafede,
  Dolag, Ettori, Fabjan, Giocoli, Mazzotta, Merten, Radovich, \&
  Tornatore}]{1367-2630-14-5-055018}
Rasia, E., Meneghetti, M., Martino, R., {et~al.} 2012, New Journal of Physics,
  14, 055018.
\newblock \url{http://stacks.iop.org/1367-2630/14/i=5/a=055018}

\bibitem[{{Rasia} {et~al.}(2012){Rasia}, {Meneghetti}, {Martino}, {Borgani},
  {Bonafede}, {Dolag}, {Ettori}, {Fabjan}, {Giocoli}, {Mazzotta}, {Merten},
  {Radovich}, \& {Tornatore}}]{2012NJPh...14e5018R}
{Rasia}, E., {Meneghetti}, M., {Martino}, R., {et~al.} 2012, New Journal of
  Physics, 14, 055018

\bibitem[{{Reiprich} {et~al.}(2009){Reiprich}, {Hudson}, {Zhang}, {Sato},
  {Ishisaki}, {Hoshino}, {Ohashi}, {Ota}, \& {Fujita}}]{2009A&A...501..899R}
{Reiprich}, T.~H., {Hudson}, D.~S., {Zhang}, Y.-Y., {et~al.} 2009, \aap, 501,
  899

\bibitem[{{Sereno} {et~al.}(2017){Sereno}, {Covone}, {Izzo}, {Ettori},
  {Coupon}, \& {Lieu}}]{2017MNRAS.472.1946S}
{Sereno}, M., {Covone}, G., {Izzo}, L., {et~al.} 2017, \mnras, 472, 1946

\bibitem[{Shaw {et~al.}(2010)Shaw, Nagai, Bhattacharya, \&
  Lau}]{0004-637X-725-2-1452}
Shaw, L.~D., Nagai, D., Bhattacharya, S., \& Lau, E.~T. 2010, The Astrophysical
  Journal, 725, 1452.
\newblock \url{http://stacks.iop.org/0004-637X/725/i=2/a=1452}

\bibitem[{Shi \& Komatsu(2014)}]{doi:10.1093/mnras/stu858}
Shi, X., \& Komatsu, E. 2014, Monthly Notices of the Royal Astronomical
  Society, 442, 521.
\newblock \url{http://dx.doi.org/10.1093/mnras/stu858}

\bibitem[{Shi {et~al.}(2016)Shi, Komatsu, Nagai, \&
  Lau}]{doi:10.1093/mnras/stv2504}
Shi, X., Komatsu, E., Nagai, D., \& Lau, E.~T. 2016, Monthly Notices of the
  Royal Astronomical Society, 455, 2936.
\newblock \url{http://dx.doi.org/10.1093/mnras/stv2504}

\bibitem[{Shi {et~al.}(2015)Shi, Komatsu, Nelson, \&
  Nagai}]{doi:10.1093/mnras/stv036}
Shi, X., Komatsu, E., Nelson, K., \& Nagai, D. 2015, Monthly Notices of the
  Royal Astronomical Society, 448, 1020.
\newblock \url{http://dx.doi.org/10.1093/mnras/stv036}

\bibitem[{Siegel {et~al.}(2018)Siegel, Sayers, Mahdavi, Donahue, Merten,
  Zitrin, Meneghetti, Umetsu, Czakon, Golwala, Postman, Koch, Koekemoer, Lin,
  Melchior, Molnar, Moustakas, Mroczkowski, Pierpaoli, \&
  Shitanishi}]{Siegel_2018}
Siegel, S.~R., Sayers, J., Mahdavi, A., {et~al.} 2018, The Astrophysical
  Journal, 861, 71.
\newblock \url{https://doi.org/10.3847%2F1538-4357%2Faac5f8}

\bibitem[{{Simet} {et~al.}(2015){Simet}, {Battaglia}, {Mandelbaum}, \&
  {Seljak}}]{2015AAS...22544304S}
{Simet}, M., {Battaglia}, N., {Mandelbaum}, R., \& {Seljak}, U. 2015, in
  American Astronomical Society Meeting Abstracts, Vol. 225, American
  Astronomical Society Meeting Abstracts \#225, 443.04

\bibitem[{Simet {et~al.}(2016)Simet, Battaglia, Mandelbaum, \&
  Seljak}]{10.1093/mnras/stw3322}
Simet, M., Battaglia, N., Mandelbaum, R., \& Seljak, U. 2016, Monthly Notices
  of the Royal Astronomical Society, 466, 3663.
\newblock \url{https://doi.org/10.1093/mnras/stw3322}

\bibitem[{{Simionescu} {et~al.}(2011){Simionescu}, {Allen}, {Mantz}, {Werner},
  {Takei}, {Morris}, {Fabian}, {Sanders}, {Nulsen}, {George}, \&
  {Taylor}}]{2011Sci...331.1576S}
{Simionescu}, A., {Allen}, S.~W., {Mantz}, A., {et~al.} 2011, Science, 331,
  1576

\bibitem[{{Smith} {et~al.}(2016){Smith}, {Mazzotta}, {Okabe}, {Ziparo},
  {Mulroy}, {Babul}, {Finoguenov}, {McCarthy}, {Lieu}, {Bah{\'e}}, {Bourdin},
  {Evrard}, {Futamase}, {Haines}, {Jauzac}, {Marrone}, {Martino}, {May},
  {Taylor}, \& {Umetsu}}]{2016MNRAS.456L..74S}
{Smith}, G.~P., {Mazzotta}, P., {Okabe}, N., {et~al.} 2016, \mnras, 456, L74

\bibitem[{{Springel} {et~al.}(2001){Springel}, {White}, {Tormen}, \&
  {Kauffmann}}]{2001MNRAS.328..726S}
{Springel}, V., {White}, S. D.~M., {Tormen}, G., \& {Kauffmann}, G. 2001,
  \mnras, 328, 726

\bibitem[{{Springel} {et~al.}(2018){Springel}, {Pakmor}, {Pillepich},
  {Weinberger}, {Nelson}, {Hernquist}, {Vogelsberger}, {Genel}, {Torrey},
  {Marinacci}, \& {Naiman}}]{Springel_et_al_2018}
{Springel}, V., {Pakmor}, R., {Pillepich}, A., {et~al.} 2018, \mnras, 475, 676

\bibitem[{{Sunyaev} \& {Zeldovich}(1970)}]{1970Ap&SS...7....3S}
{Sunyaev}, R.~A., \& {Zeldovich}, Y.~B. 1970, \apss, 7, 3

\bibitem[{{Tinker} {et~al.}(2008){Tinker}, {Kravtsov}, {Klypin}, {Abazajian},
  {Warren}, {Yepes}, {Gottl{\"o}ber}, \& {Holz}}]{2008ApJ...688..709T}
{Tinker}, J., {Kravtsov}, A.~V., {Klypin}, A., {et~al.} 2008, \apj, 688, 709

\bibitem[{Trac {et~al.}(2011)Trac, Bode, \& Ostriker}]{0004-637X-727-2-94}
Trac, H., Bode, P., \& Ostriker, J.~P. 2011, The Astrophysical Journal, 727,
  94.
\newblock \url{http://stacks.iop.org/0004-637X/727/i=2/a=94}

\bibitem[{{Urban} {et~al.}(2011){Urban}, {Werner}, {Simionescu}, {Allen}, \&
  {B{\"o}hringer}}]{2011MNRAS.414.2101U}
{Urban}, O., {Werner}, N., {Simionescu}, A., {Allen}, S.~W., \&
  {B{\"o}hringer}, H. 2011, \mnras, 414, 2101

\bibitem[{{Vikhlinin} {et~al.}(2006){Vikhlinin}, {Kravtsov}, {Forman}, {Jones},
  {Markevitch}, {Murray}, \& {Van Speybroeck}}]{2006ApJ...640..691V}
{Vikhlinin}, A., {Kravtsov}, A., {Forman}, W., {et~al.} 2006, \apj, 640, 691

\bibitem[{{Voit}(2005)}]{2005RvMP...77..207V}
{Voit}, G.~M. 2005, Reviews of Modern Physics, 77, 207

\bibitem[{{von der Linden} {et~al.}(2014){von der Linden}, {Mantz}, {Allen},
  {Applegate}, {Kelly}, {Morris}, {Wright}, {Allen}, {Burchat}, {Burke},
  {Donovan}, \& {Ebeling}}]{2014MNRAS.443.1973V}
{von der Linden}, A., {Mantz}, A., {Allen}, S.~W., {et~al.} 2014, \mnras, 443,
  1973

\bibitem[{{Zhao}(1996)}]{1996MNRAS.278..488Z}
{Zhao}, H. 1996, \mnras, 278, 488

\end{thebibliography}

\end{document}